\documentclass[review,preprint,3p,12pt,numbers,sort&compress]{elsarticle}
\usepackage{subcaption}
\usepackage{float}
\usepackage{algorithm}
\usepackage{algpseudocode}
\usepackage{latexsym}
\usepackage{placeins}
\usepackage{lineno}
\usepackage{xcolor}
\usepackage{amssymb}
\usepackage{amsmath}
\usepackage{tabularx}
\usepackage{booktabs}
\usepackage{hyperref}
\begin{document}

\begin{frontmatter}

\title{Projection-based segregated, iterative, and multi-stage physics-informed neural network for incompressible Navier–Stokes equations}

\author[1,2]{Zhenyu Zhang}
\author[1,2]{Zhixiang Ling}
\author[1,2]{Yinghao Huo}
\author[2]{Huayong Liu}
\author[2]{Yunde Su}
\ead{yundesu@scu.edu.cn}

\affiliation[1]{organization={College of Computer Science, Sichuan University},
                city={Chengdu},
                postcode={610065},
                country={China}}

\affiliation[2]{organization={National Key Laboratory of Fundamental Algorithms and Models for Engineering Simulation, Sichuan University},
                city={Chengdu},
                postcode={610207},
                country={China}}

\cortext[cor1]{Corresponding author.}
\begin{abstract}
Physics-informed neural networks\,(PINNs) offer a mesh-free and continuously differentiable approach to fluid simulation. 
However, standard monolithic PINNs for incompressible Navier-Stokes\,(NS) equations often suffer from severe constraint coupling and spectral biases, limiting both convergence and accuracy.
To address this challenge, we propose a segregated, iterative, and multi-stage PINN\,(SIM-PINN) inspired by the classical projection method in computational fluid dynamics. 
In SIM-PINN, the nonlinear NS equations are decoupled into the linearized momentum and pressure Poisson equations.
Each is addressed separately using dedicated multi-stage neural networks and coupled through projection-based inner iterations.
Each stage of the multi-stage neural networks for the momentum equations corresponds to one iteration of the projection method, in which the velocity fields are progressively refined at targeted frequencies.
Upon iteration convergence, solutions to the fully coupled and nonlinear NS equations are recovered.
By integrating the segregated and iterative framework with the multi-stage training, SIM-PINN alleviates the optimization difficulties of the incompressible NS equations arising from nonlinearity, constraint coupling, and spectral biases.
The improved convergence and accuracy are demonstrated by simulations of steady lid-driven cavity and unsteady Kolmogorov flows.
\end{abstract}


\end{frontmatter}


\section{Introduction}
\label{sec:intro}

With the recent development of deep learning, physics-informed neural networks\,(PINNs)~\cite{raissi2019physics,raissi2020hidden} have emerged as a promising alternative to classical numerical methods for solving partial differential equations\,(PDEs). 
The central idea of PINNs is to use automatic differentiation~\cite{baydin2018automatic} to compute derivatives of the neural-network output with respect to the input variables, and to incorporate physical constraints, such as governing equations, boundary conditions, and initial conditions, into the training objective in a unified manner. 
Unlike traditional numerical methods, which typically rely on mesh-based discretizations, PINNs directly construct a continuous functional approximation of the target physical field using neural networks. 
This feature provides considerable flexibility in handling complex geometries, assimilating sparse observational data, and formulating forward and inverse problems within a unified framework.

In incompressible flow simulations, while PINNs have demonstrated considerable potential~\cite{jin2021nsfnets,zhu2024physics,patel2024turbulence,
cai2024physics,xiao2024least,gao2025grid,
khademi2025physics,liu2025icpinn,wei2025ffv,liu2025automatic}, their practical application is often hindered by convergence difficulties and limited accuracy when compared with established numerical methods.
Incompressible flows are governed by the nonlinear Navier--Stokes\,(NS) equations, in which the velocity and pressure fields are intrinsically coupled and highly sensitive to boundary and initial conditions. 
Consequently, conventional monolithic PINN formulations must balance multiple loss components within a single training objective, including the governing equations, initial conditions, boundary conditions, and possibly observational data. 
Such composite losses can readily give rise to gradient-flow pathologies~\cite{wang2021understanding}, imbalanced neural tangent kernel\,(NTK) convergence rates~\cite{wang2022when}, and optimization stiffness induced by PDE regularization~\cite{krishnapriyan2021characterizing}, thereby preventing convergence to sufficiently accurate solutions.
It has also been revealed that the ill-conditioning of PINNs is closely related to the Jacobian matrix of the PDE system~\cite{cao2025analysis}.
On the other hand, the preferential learning of low-frequency modes by deep networks gives rise to frequency bias~\cite{xu2019frequency,wang2024multi} and eigenvector bias~\cite{wang2021eigenvector}, making it difficult for PINNs to accurately represent multi-scale flow structures in turbulent flows~\cite{zhang2025physics,cao2025rans,wang2025simulating}.

To address the convergence and accuracy bottlenecks of PINNs, extensive studies have been conducted across various directions, including loss adaptation, physical-constraint imposition, geometric and feature preprocessing, advanced optimization strategies, and temporal causality preservation.
In terms of loss adaptation, gradient-statistics balancing~\cite{wang2021understanding}, NTK-based adaptive weighting~\cite{wang2022when}, and self-adaptive weighting~\cite{mcclenny2023self} have been developed to rebalance different loss components.
Additionally, gradient-enhanced PINNs~\cite{yu2022gradient} incorporate the gradient of the PDE residual into the loss function to enhance training convergence and solution accuracy.
Regarding physical-constraint imposition, embedding physical priors directly into network representations---such as strictly enforcing boundary conditions via distance functions~\cite{sukumar2022exact} or incorporating inherent symmetries through adaptive symmetry architectures~\cite{liu2024aspinn}---effectively reduces the optimization burden for complex physical fields.
In the context of geometric and feature preprocessing, approaches including the variable linear transformation~\cite{wu2024variable}, coordinate transformations~\cite{zhang2025physics}, and feature-enhanced neural networks~\cite{song2025fenn} have been shown to alleviate learning difficulties for complex flows. 
From an optimization perspective, techniques such as the Nystr{\"o}m-Newton conjugate gradient method~\cite{rathore2024challenges}, modified BFGS methods~\cite{urban2025unveiling}, matrix preconditioning~\cite{song2025matrix}, and natural gradient methods~\cite{jnini2025gauss} have been proposed to improve conditioning and the quality of the search direction. 
Meanwhile, adaptive transfer learning along minimum energy paths~\cite{liu2023adaptive} enhances training stability across complex loss landscapes.
Finally, concerning temporal causality preservation, causal time weighting~\cite{wang2024respecting}, sequential time segmentation~\cite{krishnapriyan2021characterizing}, and pseudo-time-marching~\cite{cao2025analysis,cao2025rans} have been shown to enhance the performance of PINNs.

Despite these advances, it remains challenging for PINNs to accurately and stably simulate incompressible flows at relatively high Reynolds numbers~\cite{wang2025simulating,zhang2026physics}, where resolving multi-scale and potentially unsteady flow structures is essential.
Most existing studies rely on a monolithic framework, in which the strong coupling of nonlinear equations with distinct mathematical characteristics exacerbates optimization stiffness and gradient pathologies. This challenge motivates the development of new PINN architectures tailored to decouple the governing equations rather than solving the system monolithically. 

In this work, we propose a segregated, iterative, and multi-stage PINN\,(SIM-PINN) framework for incompressible flow simulations. 
This framework is inspired by the projection method in classical low-Mach-number computational fluid dynamics\,(CFD)~\cite{chorin1968numerical}.
Unlike conventional monolithic PINNs, the SIM-PINN architecture decomposes the coupled governing system into structured subproblems, such as linearized momentum prediction and pressure correction, each handled by dedicated neural networks. 
This segregated approach avoids direct gradient competition between the momentum and continuity equations within a single loss function and is designed to reduce the optimization burden through linearization.
The outputs of these sub-networks are reconciled through projection-based updates and coupled through inner iterations, driving the segregated solutions toward physical consistency while limiting error accumulation during decoupled training.
In addition, a multi-stage correction strategy inspired by Wang and Lai~\cite{wang2024multi} is introduced within SIM-PINN to mitigate the spectral bias of deep networks. 
This stage-wise refinement enables the model to progressively resolve higher-frequency residual components that are difficult for a single network to learn.
By combining segregated decomposition, iterative coupling, and stage-wise refinement, the proposed SIM-PINN provides a structured approach to alleviate loss conflicts, reduce optimization stiffness, and improve the resolution of complex structures in incompressible flow simulations.

The remainder of this paper is organized as follows.
Section~\ref{sec:governing_eqn} introduces the governing equations for incompressible flows.
Section~\ref{sec:SIM-PINN} details the proposed SIM-PINN framework, in which the projection method in classical CFD is briefly reviewed, followed by a detailed description of the SIM-PINN algorithm 
and a discussion of the multi-stage neural networks within the framework.
Numerical results for steady and unsteady flow problems are presented and discussed in Section~\ref{sec:results}.
\textcolor{black}{Section~\ref{sec:discussions} discusses the main contributions, computational considerations, and future perspectives of the proposed framework.}
Section~\ref{sec:conclusions} summarizes the main conclusions of this work.

\section{Governing Equations}
\label{sec:governing_eqn}
The dynamics of incompressible flows are governed by the NS equations, which can be expressed as:
\begin{equation}
\frac{\partial u_i}{\partial t}+u_j\frac{\partial u_i}{\partial x_j} + \frac{1}{\rho}\frac{\partial p}{\partial x_i} - \nu \frac{\partial^2 u_i}{\partial x_j \partial x_j}-\frac{1}{\rho}f_i=0,
\label{eq:momen}
\end{equation}
\begin{equation}
\frac{\partial u_j}{\partial x_j} = 0,
\label{eq:mass}
\end{equation}
where $u_i$ denotes the velocity components, $p$ represents the pressure, and $\nu$ is the kinematic viscosity.
Additionally, $\rho$ and $f_i$ denote the density and the body force per unit volume, respectively.
Here, the Einstein summation convention over repeated indices is applied in Eqs.~\eqref{eq:momen} and~\eqref{eq:mass}.
For steady incompressible flows, the temporal derivative term in Eq.~\eqref{eq:momen} naturally vanishes.

\section{SIM-PINN: Segregated, Iterative, and Multi-Stage Physics-Informed Neural Network}
\label{sec:SIM-PINN}

\subsection{Projection-based segregated iterative method}
\label{sec:projection_cfd}
The governing equations for the incompressible flows have a mixed parabolic-elliptic character.
In traditional CFD, these equations are typically solved using the projection-based segregated iterative method, which has demonstrated sound stability and accuracy in numerous studies~\cite{chorin1968numerical,kim1985application,brown2001accurate,guermond2006overview,Ferziger2002book,Desjardins2008,Su2018,kim2020front,su2023interaction}.
The essential idea is to employ the Helmholtz--Hodge decomposition, by which a general intermediate velocity field is decomposed into a divergence-free component and a gradient field component~\cite{chorin1968numerical,guermond2006overview}. 
For the intermediate velocity $\mathbf{u}^{*}$ obtained from the predicted momentum equation, the incompressibility constraint is not yet strictly enforced, and therefore one generally has $\nabla\cdot\mathbf{u}^{*}\neq 0$. 
A pressure correction is thus required to project the intermediate velocity onto the divergence-free velocity space.
This prediction-correction procedure is repeated iteratively so that the splitting error can be removed and the solutions to the fully coupled and non-linear governing equations can be recovered.

\begin{algorithm}[!ht]
\caption{Segregated iterative algorithm for semi-discrete NS equations}
\label{alg:cfd_algorithm_1}
\begin{algorithmic}[1]
\makeatletter
\setcounter{ALG@line}{-1}
\makeatother
\State Given the converged solutions at the previous time step $t^n$, initialize the pressure and velocities for the forthcoming iterative procedure:
\begin{equation*}
p_0=p, \quad u^{n+1}_{i,0} = u_i^{n}.
\end{equation*}

\Statex \textbf{For} the sub-iteration $k=0, 1, \dots, Q-1$:
\State  Solve the momentum equation to obtain the intermediate velocity $u^{*,n+1}_{i,k+1}$. 
The form of the semi-discrete momentum equation depends on the adopted temporal scheme.
As for the first-order backward Euler scheme, it is formulated as,
\begin{equation}
\frac{u_{i,k+1}^{*,n+1}-u_i^{n}}{\Delta t}+u_{j,k}^{n+1}\frac{\partial u_{i,k+1}^{*,n+1}}{\partial x_j} + \frac{1}{\rho}\frac{\partial p_{k}}{\partial x_i} - \nu \frac{\partial ^2 u_{i,k+1}^{*,n+1}}{\partial x_j \partial x_j} - \frac{1}{\rho}f_i=0. 
\label{eq:alg_momen_Euler}
\end{equation}
And for the second-order Crank--Nicolson scheme, it is expressed as,
\begin{equation}
\frac{u_{i,k+1}^{*,n+1}-u_{i}^{n}}{\Delta t} + \frac{u_j^{n}+u_{j,k}^{n+1}}{2} \frac{\partial \left[\left(u_i^{n}+u_{i,k+1}^{*,n+1}\right)/2\right]}{\partial x_j} + \frac{1}{\rho}\frac{\partial p_k}{\partial x_i} - \nu \frac{\partial^2 \left(u_i^{n}+u_{i,k+1}^{*,n+1}\right)/2}{\partial x_j \partial x_j} - \frac{1}{\rho}f_i=0.
\label{eq:alg_momen_CN}
\end{equation}
\State  Solve the Poisson equation to obtain the pressure correction $\delta p$:
\begin{equation}
\frac{\partial ^2 \delta p}{\partial x_j \partial x_j}=\frac{\rho}{\Delta t}\frac{\partial u^{*,n+1}_{j,k+1}}{\partial x_j}. 
\label{eq:poisson}
\end{equation}
\State  Correct the velocity and update the pressure:
\begin{equation}
u^{n+1}_{i,k+1} = u^{*,n+1}_{i,k+1} - \frac{\Delta t}{\rho} \frac{\partial \delta p}{\partial x_i}, \quad
p_{k+1} = p_{k} + \delta p.
\label{eq:pressure_correct}
\end{equation}
\State  Increment $k = k + 1$, and proceed to the next sub-iteration.
\State Upon convergence of the sub-iterations (or reaching $k=Q$), advance to the next time step $t^{n+1}$ by updating the solutions:
\begin{equation}
u^{n+1}_i = u^{n+1}_{i,k}, \quad p = p_k.
\label{eq:update_sol}
\end{equation}
\end{algorithmic}
\end{algorithm}
In the present work, two different time-stepping algorithms are considered.
\textcolor{black}{For the steady flow simulation with the pseudo-time marching approach, the first-order backward Euler scheme is employed.
}
For the unsteady flow simulation with higher temporal accuracy requirement, the second-order accurate Crank--Nicolson scheme is adopted.
The segregated iterative algorithm for the semi-discrete NS equations with the two different time-stepping method is presented in Algorithm~\ref{alg:cfd_algorithm_1}.
Since mathematically, the pressure in Eq.~\eqref{eq:momen} serves as a Lagrange multiplier to enforce continuity, the superscript information has been omitted for the pressure in Algorithm~\ref{alg:cfd_algorithm_1} for brevity.
The body force $f_i$ in Algorithm~\ref{alg:cfd_algorithm_1} is assumed to be time-invariant.
In each sub-iteration, the momentum equation\,(Eqs.~\eqref{eq:alg_momen_Euler} and~\eqref{eq:alg_momen_CN}) and the Poisson equation\,(Eq.~\eqref{eq:poisson}) are linear, which is important for the SIM-PINN presented in Section~\ref{sec:algorithm_simpinn}.

\subsection{Overview of SIM-PINN}
\label{sec:algorithm_simpinn}

\begin{figure} 
    \centering
    \includegraphics[width=\textwidth]{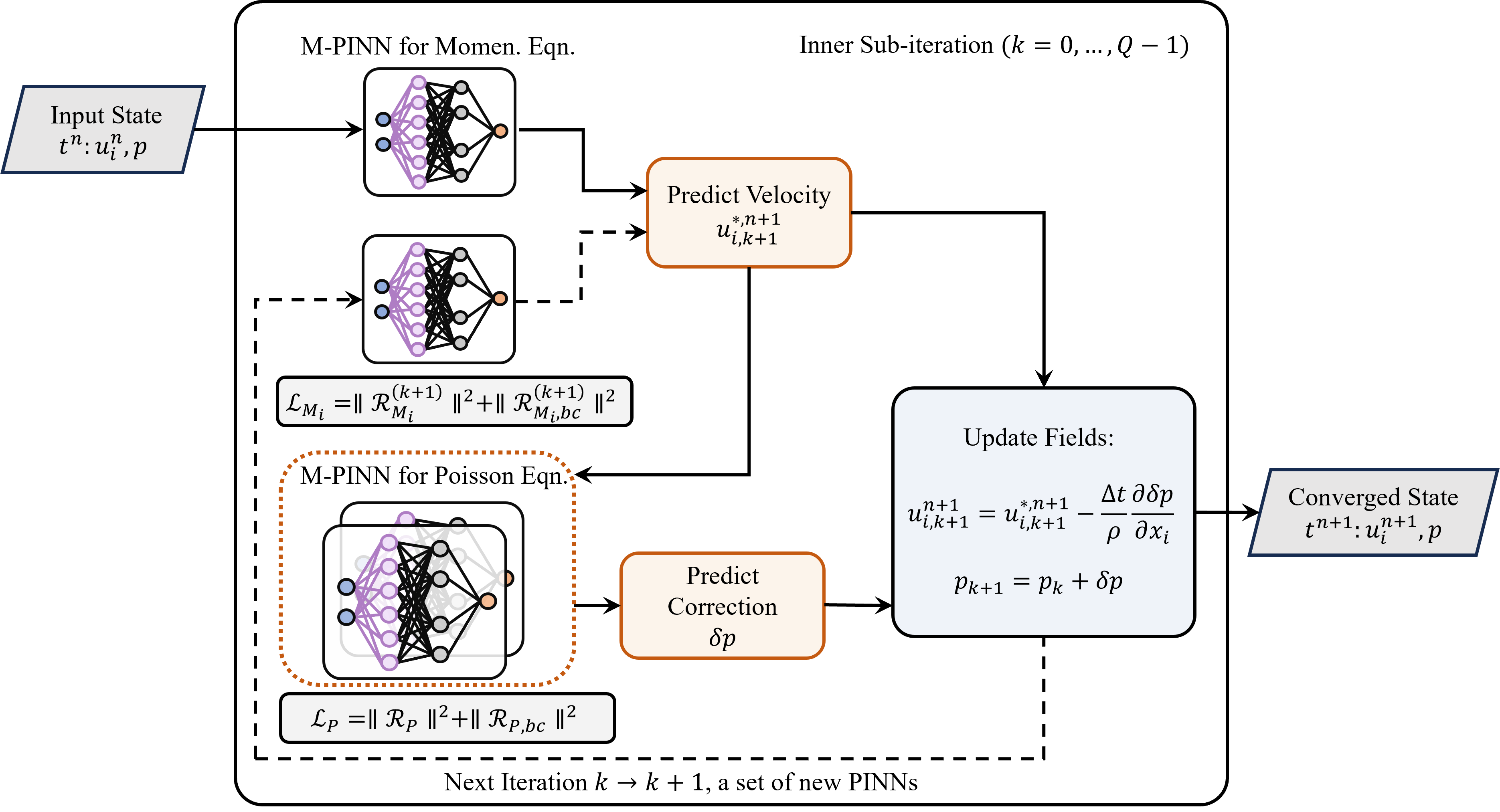}
    \caption{Schematic illustration of the SIM-PINN framework.}
    \label{fig:SIM_schematic}
\end{figure}

The proposed SIM-PINN algorithm is inspired by the projection-based segregated iterative method as described in Algorithm~\ref{alg:cfd_algorithm_1}.
The schematic of SIM-PINN is illustrated in Fig.~\ref{fig:SIM_schematic}.
Within each time step, the velocity field is first updated based on the momentum equations, and is then corrected through a pressure-correction step so that the incompressibility constraint $\partial u_j/{\partial x_j}=0$ is satisfied. 
Unlike traditional numerical methods, SIM-PINN represents the momentum-prediction and pressure-correction sub-problems by distinct neural networks, respectively.
The coupling of these neural networks is accomplished through inner sub-iterations.
Each sub-iteration of the momentum equation corresponds to one stage in the multi-stage neural-network framework.
Instead of predicting the velocity field from the momentum equation directly, the neural network predicts the velocity change at each stage.
In this way, the optimization task is formulated as learning a corrective update on top of the existing velocity field, which is essential for the success of SIM-PINN.

In what follows, the SIM-PINN algorithm is described by taking a single time step as an example. 
With a uniform time step size $\Delta t$, the velocity and pressure fields are advanced from time $t^n$ to $t^{n+1}$. 

\begin{algorithmic}[1]
\makeatletter
\setcounter{ALG@line}{-1}
\makeatother
\State Given the converged solutions at the previous time step $t^n$, initialize the pressure and velocities for the forthcoming iterative procedure:
\begin{equation*}
p_0=p, \quad u^{n+1}_{i,0} = u_i^{n}.
\end{equation*}

\Statex \textbf{For} the sub-iteration $k=0, 1, \dots, Q-1$:
\State  Predict the velocity correction, $\delta u_{i,k+1}^{*}$,  to obtain the intermediate velocity, $u^{*,n+1}_{i,k+1}$, with a new neural network $\mathcal{N}_{{M}_i}^{(k+1)}$:
\begin{equation}
\widehat{u}_{i,{k+1}}=\mathcal{N}_{{M}_i}^{(k+1)}(\mathbf{x}),
\qquad
\delta u_{i,k+1}^{*}=\epsilon_{M_i}^{(k+1)}\widehat{u}_{i,k+1},
\qquad
u^{*,n+1}_{i,k+1} = u^{*,n+1}_{i,k} + \delta u_{i,k+1}^{*},
\qquad
\mathbf{x} \in X_\Omega .
\label{eq:ui_iteration}
\end{equation}
Here, $\Omega$ denotes the computational domain.
$\epsilon_{M_i}^{(k+1)}$ is a magnitude prefactor in the multi-stage neural network and is to be discussed and evaluated in Section~\ref{sec:mpinn-momen}.
$u^{*,n+1}_{i,k}=0$ when $k=0$.
Same as in Algorithm~\ref{alg:cfd_algorithm_1}, the residual of the momentum equation, $\mathcal{R}_{M_i}^{(k+1)}$, depends on the adopted temporal scheme.
As for the first order backward Euler scheme, it is formulated as,
\begin{equation}
\mathcal{R}_{M_i}^{(k+1)} = \frac{u_{i,k+1}^{*,n+1}-u_i^{n}}{\Delta t}+u_{j,k}^{n+1}\frac{\partial u_{i,k+1}^{*,n+1}}{\partial x_j} + \frac{1}{\rho}\frac{\partial p_{k}}{\partial x_i} - \nu \frac{\partial ^2 u_{i,k+1}^{*,n+1}}{\partial x_j \partial x_j} - \frac{1}{\rho}f_i. 
\label{eq:simpinn_momen_Euler}
\end{equation}
And for the second-order Crank--Nicolson scheme, it is expressed as,
\begin{equation}
\mathcal{R}_{M_i}^{(k+1)} = \frac{u_{i,k+1}^{*,n+1}-u_{i}^{n}}{\Delta t} + \frac{u_j^{n}+u_{j,k}^{n+1}}{2} \frac{\partial \left[\left(u_i^{n}+u_{i,k+1}^{*,n+1}\right)/2\right]}{\partial x_j} + \frac{1}{\rho}\frac{\partial p_k}{\partial x_i} - \nu \frac{\partial^2 \left(u_i^{n}+u_{i,k+1}^{*,n+1}\right)/2}{\partial x_j \partial x_j} - \frac{1}{\rho}f_i.
\label{eq:simpinn_momen_CN}
\end{equation}
The boundary residual is defined as,
\begin{equation}
\mathcal{R}_{M_i,bc}^{(k+1)}(\mathbf{x}_b)
=
u_{i,k+1}^{*,n+1}(\mathbf{x}_b)-u_{i,bc}(\mathbf{x}_b),
\quad
\mathbf{x}_b\in X_{\partial\Omega},
\label{eq:BC_iteration}
\end{equation}
where $\partial\Omega$ and $u_{i,bc}$ denote the boundary of the computational domain and the velocity values at the boundary, respectively.
The total loss for the momentum equation in the $i$ direction is written as,
\begin{equation}
\mathcal{L}_{M_i}(\theta)
=
\frac{1}{N_\Omega}
\sum_{j=1}^{N_{\Omega}}
\left\|
\mathcal{R}_{M_i}^{(k+1)}(\mathbf{x}^{j})
\right\|_2^2
+
\frac{1}{N_{\partial\Omega}}
\sum_{j=1}^{N_{\partial\Omega}}
\left\|
\mathcal{R}_{M_i,bc}^{(k+1)}(\mathbf{x}_b^j)
\right\|_2^2,
\label{eq:LOSS_mom}
\end{equation}
where $\{\mathbf{x}^j\}_{j=1}^{N_{\Omega}}$ and $\{\mathbf{x}_b^j\}_{j=1}^{N_{\partial\Omega}}$ are the collocation points in the computation domain and the boundary, respectively.
\State Solve the Poisson equation with the multi-stage PINN to obtain the pressure correction $\delta p$.
The corresponding residual of the Poisson equation is expressed as,
\begin{equation}
\mathcal{R}_{P}(\mathbf{x})
=
\frac{\partial ^2 \delta p(\mathbf{x})}{\partial x_j \partial x_j}
-
\frac{\rho}{\Delta t}\frac{\partial u_{j,k+1}^{*,n+1}}{\partial x_j}.
\label{eq:poisson_k}
\end{equation}
With the Neumann boundary condition, the boundary residual is defined as,
\begin{equation}
\mathcal{R}_{P,bc}(\mathbf{x}_b)
=
n_i\frac{\partial \delta p(\mathbf{x}_b)}{\partial x_i},
\quad
\mathbf{x}_b\in X_{\partial\Omega},
\label{eq:poisson_bc_residual}
\end{equation}
where $n_i$ denotes the $i$-th component of the unit normal vector on the boundary $\partial\Omega$.
The total loss function for the pressure-correction Poisson equation is written as
\begin{equation}
\mathcal{L}_{P}(\theta)
=
\frac{1}{N_\Omega}
\sum_{j=1}^{N_\Omega}
\left\|
\mathcal{R}_{P}(\mathbf{x}^j)
\right\|_2^2
+
\frac{1}{N_{\partial\Omega}}
\sum_{j=1}^{N_{\partial\Omega}}
\left\|
\mathcal{R}_{P,bc}(\mathbf{x}_b^j)
\right\|_2^2 .
\label{eq:LSS_poisson}
\end{equation}
The details of the multi-stage PINN for the Poisson equation is to be presented in Section~\ref{sec:mpinn-Poisson}.

\State Correct the accumulated velocity field and update the pressure field with Eq.~\eqref{eq:pressure_correct}.

\State  Increment $k = k + 1$, and proceed to the next sub-iteration.
\State Upon convergence of the sub-iterations (or reaching $k=Q$), advance to the next time step $t^{n+1}$ after updating the solutions using Eq.~\eqref{eq:update_sol}.
\end{algorithmic}

\subsection{Multi-stage neural networks in SIM-PINN}
\label{sec:mpinn_sim-pinn}
In the SIM-PINN, the multi-stage training is essential for the deep convergence and stability of the framework.
The concept of the multi-stage neural networks is reflected in both the solution of the Poisson equation and the momentum equations.
In the Poisson equation, which is a second-order linear equation, the multi-stage neural network is a standalone solver.
In contrast, in solving the momentum equation, the multi-stage concept is coupled with the iterative sub-steps of the projection method.
Rather than solving the momentum equation using a single network, a new network is introduced at each sub-iteration to learn the corrections unresolved in the preceding stages, before the continuity enforcement.
As shown in Eq.~\eqref{eq:simpinn_momen_Euler}, at each sub-iteration within the segregated and iterative algorithm, the momentum equation has been linearized, which favors the application of the multi-stage training to the solution of the non-linear NS equations.

\subsubsection{Multi-stage neural networks}
\label{sec:mpinn}
\FloatBarrier
The multi-stage neural network framework~\cite{wang2024multi} sequentially trains multiple sub-networks to iteratively resolve the high-frequency residuals of governing equations. 
Consider a general PDE defined by $\mathcal{F}[u(\mathbf{x})] = 0$. 
The multi-stage training procedure can be unified for any stage $k \ge 1$ by defining an initial zero state $u^{(0)}(\mathbf{x}) \equiv 0$.
At the $k$-th stage, a sub-network is introduced to compensate for the approximation error left by the previous stages. 
The updated global prediction $u^{(k)}(\mathbf{x})$ is constructed by superposition:
\begin{equation}
u^{(k)}(\mathbf{x}) = u^{(k-1)}(\mathbf{x}) +  \epsilon^{(k)} \mathcal{N}^{(k)} (\mathbf{x}; \kappa_k, \boldsymbol{\theta}^{(k)}),
    \label{eq:msnn_superposition}
\end{equation}
where $\mathcal{N}^{(k)}$ represents a standard multi-layer perceptron neural network, $\epsilon^{(k)}$ is a magnitude prefactor, $\kappa_k$ is a scale factor, and $\boldsymbol{\theta}^{(k)}$ denotes parameters of $\mathcal{N}^{(k)}$.
During the $k$-th stage training, the parameters of all previously trained networks ($\boldsymbol{\theta}^{(1)}, \dots, \boldsymbol{\theta}^{(k-1)}$) are strictly frozen. 
The optimization is performed solely on $\boldsymbol{\theta}^{(k)}$ by minimizing the total loss function of the PDE and boundary conditions.

For the initial stage\,($k=1$), the method naturally recovers a standard PINN by setting $\epsilon^{(1)} = 1$ and $\kappa_1 = 1$. 
For higher stages\,($k \ge 2$), $\epsilon^{(k)}$ rescales the network output 
to normalize the increasingly small target residual to $\mathcal{O}(1)$. 
And the architecture of $\mathcal{N}^{(k)}$ is further tailored by multiplying 
the scale factor $\kappa_k$, proportional to the dominant frequency of the residual in the preceding stage, to the weights of the first hidden layer with a sinusoidal 
activation function.
The combination of the staged training and the specific design of each new-stage network mitigates the  spectral biases of conventional neural networks, enabling remarkable accuracy.

The standard multi-stage neural network is difficult to apply effectively to the nonlinear and strongly coupled NS equation systems involving multiple interacting fields~\cite{wang2025simulating}.
Within the SIM-PINN framework, due to the projection-based iterative decoupling, we are able to extend the multi-stage network concept to the full NS equations.
In the following subsections, the multi-stage networks for the Poisson and momentum equations are presented, with a focus on the specific parameterizations and network structures.

\subsubsection{Multi-stage neural network for the Poisson equation}
\label{sec:mpinn-Poisson}

\begin{figure}
    \centering
    \begin{subfigure}[b]{0.45\textwidth}
        \centering
        \includegraphics[width=\textwidth]{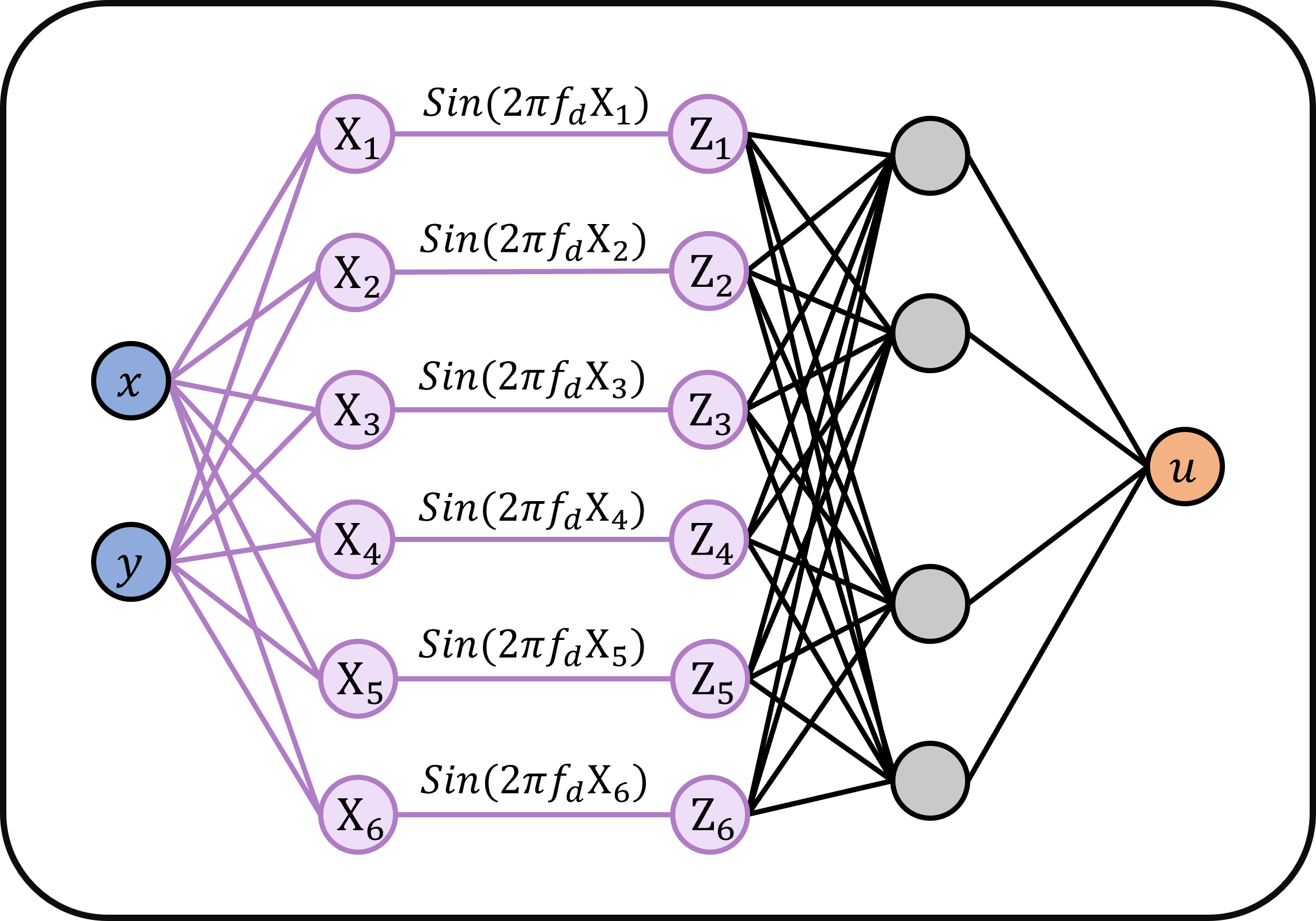}
        \caption{Sinusoidal feature embedding}
        \label{fig:embed_sin}
    \end{subfigure}
    \hfill
     \begin{subfigure}[b]{0.45\textwidth}
        \centering
        \includegraphics[width=\textwidth]{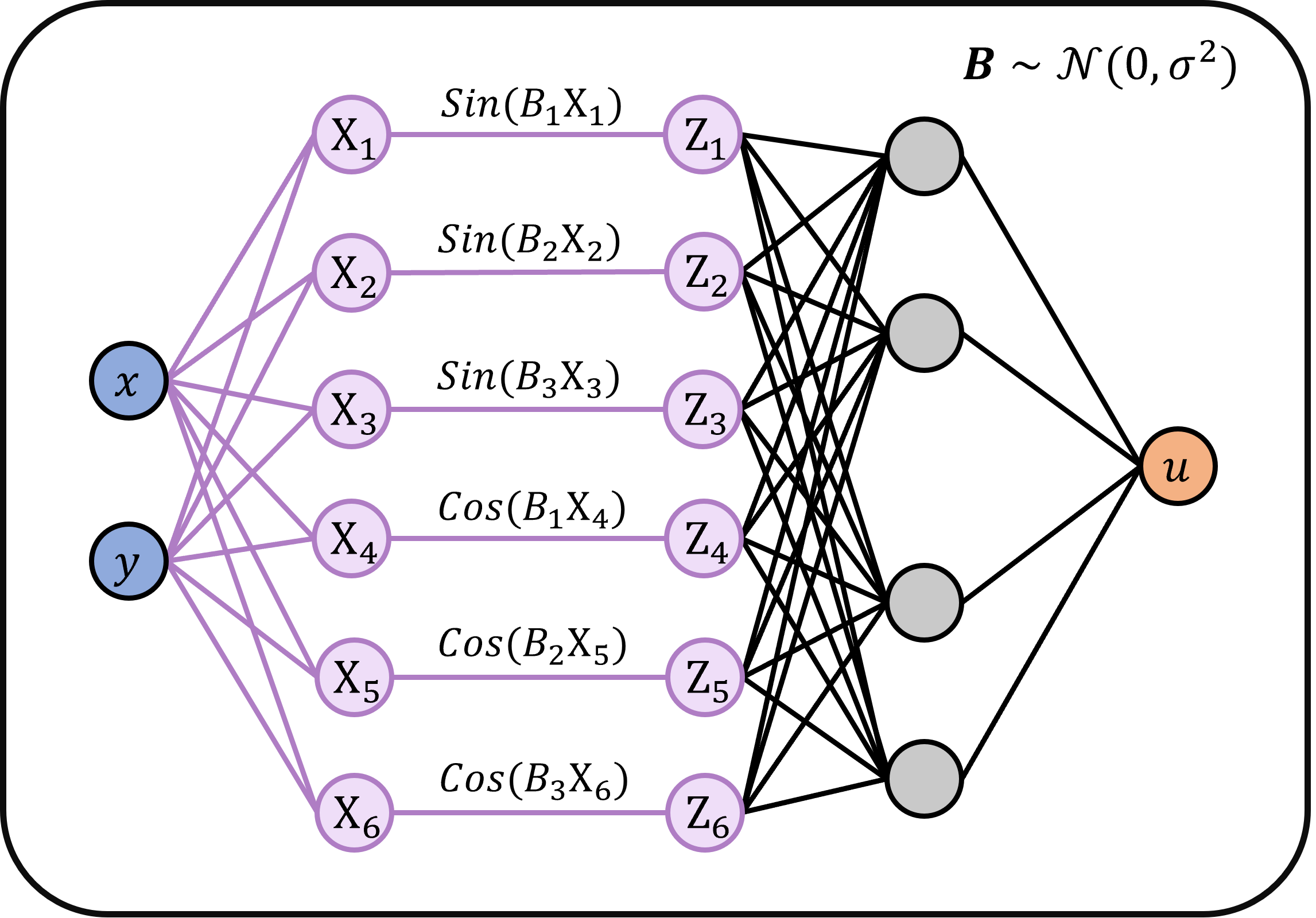}
        \caption{Random Fourier feature embedding}
        \label{fig:embed_rff}
    \end{subfigure}
    \caption{Schematic illustration of two network architectures with different feature embeddings.}
    \label{fig:network_structure}
\end{figure}

The pressure-correction Poisson equation is solved at each sub-iteration of SIM-PINN by an independent multi-stage neural-network solver. 
When solving this equation, the intermediate velocity field is fixed, and the source term of the Poisson equation is constructed from the divergence of this velocity field.

At the $k$-th stage of the multi-stage neural network for the Poisson equation, the output of the network is denoted by $\delta p_k(\mathbf{x})$. 
The accumulated pressure correction after $k$ stages is expressed as,
\begin{equation}
\delta p^{(k)}(\mathbf{x})
=
\delta p^{(k-1)}(\mathbf{x}) +\epsilon_{P}^{(k)} \delta p_k(\mathbf{x}),
\label{eq:poisson_total_output}
\end{equation}
where $\epsilon_P^{(k)}$ is the magnitude prefactor of the $k$-th stage. 
The corresponding Poisson residual is updated by
\begin{equation}
\mathcal{R}_{P}^{(k)}(\mathbf{x})
=
\mathcal{R}_{P}^{(k-1)}(\mathbf{x})
+
\epsilon_P^{(k)}
\nabla^2 \delta p_k(\mathbf{x}) .
\label{eq:R_PDE_Poisson_iteration}
\end{equation}
The sign in Eq.~\eqref{eq:R_PDE_Poisson_iteration} follows the residual definition used in Eq.~\eqref{eq:poisson_k}.

Two feature-embedding network structures are employed in the Poisson solver, as illustrated in Fig.~\ref{fig:network_structure}: the random Fourier feature\,(RFF) embedding~\cite{rahimi2007random,tancik2020fourier} and the frequency-guided sinusoidal embedding~\cite{wang2024multi}. 
In both embedding structures, the physical coordinates are first mapped into a frequency-enriched feature space before being passed to the corresponding neural network. The RFF embedding introduces a collection of randomly sampled Fourier modes spanning a broad range of spatial frequencies, whereas the frequency-guided sinusoidal embedding is constructed to represent a specific dominant frequency. Accordingly, the RFF embedding is employed in the first stage of the Poisson solver, where the residual generally exhibits a broad frequency spectrum. In subsequent stages, the frequency-guided sinusoidal embedding is adopted because the spectral energy of the remaining residual becomes progressively concentrated within narrower and higher-frequency bands.

The first-stage network output, $\delta p_1(\mathbf{x})$, is expressed as,
\begin{equation}
\delta p_1(\mathbf{x})
=
\mathcal{N}_{P,\mathrm{RFF}}
\left(
\phi_{\mathrm{RFF}}(\mathbf{x});
\boldsymbol{\theta}_{P}^{(1)}
\right),
\label{eq:poisson_rff_stage}
\end{equation}
where $\boldsymbol{\theta}_{P}^{(1)}$ denotes the trainable parameters of the first-stage neural network and $\phi_{\mathrm{RFF}}(\mathbf{x})$ represents the RFF mapping. 
The RFF mapping is defined as,
\begin{equation}
\phi_{\mathrm{RFF}}(\mathbf{x})
=
\left[
\sin\left(2\pi B\mathbf{x}\right),\,
\cos\left(2\pi B\mathbf{x}\right)
\right],
\label{eq:poisson_rff_mapping}
\end{equation}
where $B$ is a randomly sampled matrix with dimension $M\times d$.
Here, $d$ denotes the spatial dimension and $M$ is the number of randomly sampled Fourier modes. 
Each element $b_{ij}$ of the matrix $B$ is sampled from the Gaussian distribution: $b_{ij} \sim
N\left(0,\sigma^2\right)$, where $\sigma$ is the standard deviation that controls the characteristic frequency scale and spectral bandwidth of the generated Fourier features. 
In the present work, $\sigma=2.0$ is used. 
The matrix $B$ is sampled once before training and remains fixed throughout the optimization, while only the network parameters $\boldsymbol{\theta}_{P}^{(1)}$ are updated.

For subsequent stages with $k\geq2$, the network output is expressed as,
\begin{equation}
\delta p_k(\mathbf{x})
=
\mathcal{N}_{P,\mathrm{sin}}
\left(
\phi_{\mathrm{sin}}^{(k)}(\mathbf{x});
\boldsymbol{\theta}_{P}^{(k)}
\right),
\label{eq:poisson_sin_stage}
\end{equation}
where $\boldsymbol{\theta}_{P}^{(k)}$ denotes the trainable parameters of the corresponding neural network.
The frequency-guided sinusoidal mapping, $\phi_{\mathrm{sin}}^{(k)}(\mathbf{x})$, in Eq.~\eqref{eq:poisson_sin_stage} is defined as,
\begin{equation}
\phi_{\mathrm{sin}}^{(k)}(\mathbf{x})
=
\sin\left(
\kappa_P^{(k)}\mathbf{x}
\right),
\label{eq:poisson_sin_mapping}
\end{equation}
where $\kappa_P^{(k)}$ is the scale factor.
$\kappa_P^{(k)}$ is estimated using the dominant frequency, $f_{d,P}^{(k-1)}$, as 
\begin{equation}
\kappa_P^{(k)}
=
2\pi f_{d,P}^{(k-1)}.
\label{eq:poisson_kappa_fd}
\end{equation}

The dominant frequency $f_{d,P}^{(k-1)}$ is extracted from the Poisson residual of the preceding stage using a spectral-analysis procedure. 
Specifically, the residual field is first sampled on the computational grid, with its mean value removed to suppress the zero-frequency component. 
A Hamming window is subsequently applied to suppress the spectral leakage caused by the finite spatial domain. 
The residual energy spectrum is then evaluated using the discrete Fourier transform.
The dominant frequency is identified from the principal peak of this spectrum.

The magnitude prefactor $\epsilon_P^{(k)}$ is estimated from the magnitude of the residual and $f_{d,P}^{(k-1)}$ as,
\begin{equation}
\epsilon_P^{(k)}
=
\frac{
\mathrm{RMS}\left(\mathcal{R}_P^{(k-1)}\right)
}{
\left(2\pi f_{d,{P}}^{(k-1)}\right)^2
},
\label{eq:epsilon_n}
\end{equation}
where $\mathrm{RMS}\left(\mathcal{R}_{P}^{(k-1)}\right)$ denotes the root-mean-square\,(RMS) magnitude of the Poisson residual evaluated over all collocation points at the end of the preceding stage.

\subsubsection{Multi-stage neural network for the momentum equations}
\label{sec:mpinn-momen}
\begin{figure} 
    \centering
    \includegraphics[width=\textwidth]{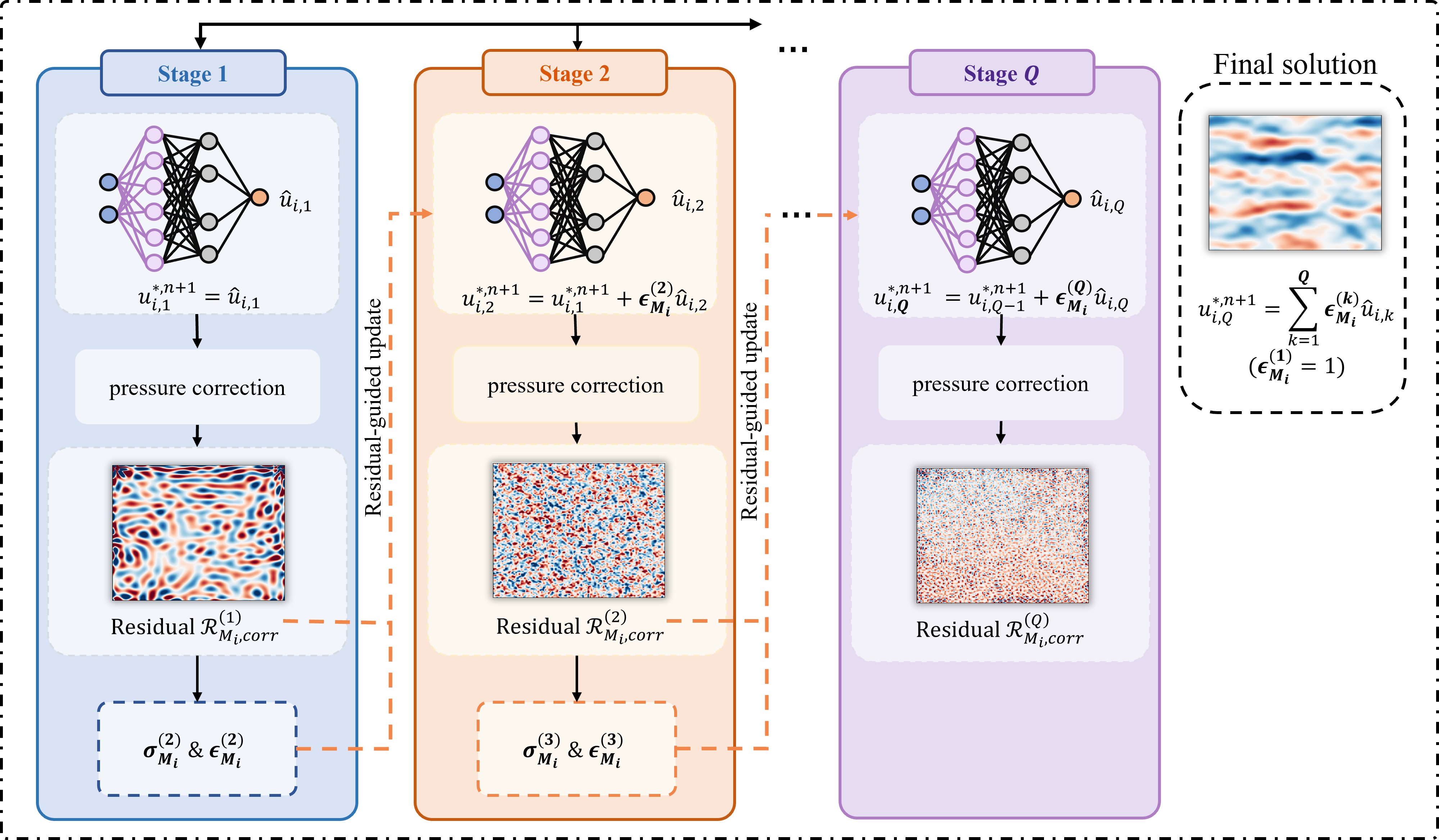}
\caption{Schematic illustration of the multi-stage neural network architecture for the momentum equation in the $i$-th coordinate direction and its iterative coupling with pressure correction.}
    \label{fig:Schematic_for_mom_u}
\end{figure}

The multi-stage neural network for each momentum equation is tightly coupled with the iterative procedure of the proposed SIM-PINN framework.
As illustrated in Fig.~\ref{fig:Schematic_for_mom_u}, each stage corresponds to one iteration.
Within each iteration, the velocity field is first updated via a dedicated neural network and subsequently corrected through the pressure Poisson equation.
The corrected residual of the momentum equation after the Poisson correction drives the training of the next-stage neural network.

At the ($k$+1)-th stage\,(iteration), the residual of the momentum equation, $\mathcal{R}_{M_i}^{(k+1)}$, can be re-formulated by substituting the stage-wise velocity correction defined by Eq.~\eqref{eq:ui_iteration} into the equation.
For the first-order Euler temporal discretization, it is expressed as,
\begin{equation}
\begin{aligned}
\mathcal{R}_{M_i}^{(k+1)}
=
\epsilon_{M_i}^{(k+1)}
\left(
\frac{\hat{u}_{i,k+1}}{\Delta t}
+
u_{j,k}^{n+1}
\frac{\partial\hat{u}_{i,k+1}}{\partial x_{j}}
-
\nu
\frac{\partial^{2}\hat{u}_{i,k+1}}{\partial x_{j}\partial x_{j}}
\right)
+
\mathcal{R}_{M_i,\mathrm{corr}}^{(k)},
\end{aligned}
\label{eq:mom_iteration_Euler_k1}
\end{equation}
where $\mathcal{R}_{M_i,\mathrm{corr}}^{(k)}$ is a corrected residual after the Poisson correction.
It is written as,
\begin{equation}
\mathcal{R}_{M_i,\mathrm{corr}}^{(k)}
=
\frac{u_{i,k}^{*,n+1}-u_i^{n}}{\Delta t}
+
u_{j,k}^{n+1}
\frac{\partial u_{i,k}^{*,n+1}}{\partial x_j}
+
\frac{1}{\rho}\frac{\partial p_{k}}{\partial x_i}
-
\nu
\frac{\partial ^2 u_{i,k}^{*,n+1}}{\partial x_j \partial x_j}
-
\frac{1}{\rho}f_i.
\label{eq:mom_iteration_Euler_corr}
\end{equation}

For the second-order Crank--Nicolson discretization, the same decomposition leads to
\begin{equation}
\begin{aligned}
\mathcal{R}_{M_i}^{(k+1)}
&=
\epsilon_{M_i}^{(k+1)}
\left(
\frac{\hat{u}_{i,k+1}}{\Delta t}
+
\frac{u_{j}^{n}+u_{j,k}^{n+1}}{4}
\frac{\partial\hat{u}_{i,k+1}}{\partial x_{j}}
-
\frac{\nu}{2}
\frac{\partial^{2}\hat{u}_{i,k+1}}{\partial x_{j}\partial x_{j}}
\right)
+
\mathcal{R}_{M_i,\mathrm{corr}}^{(k)} ,
\end{aligned}
\label{eq:momen_CN_k1}
\end{equation}
where
\begin{equation}
\mathcal{R}_{M_i,\mathrm{corr}}^{(k)}
=
\frac{u_{i,k}^{*,n+1}-u_{i}^{n}}{\Delta t}
+
\frac{u_j^{n}+u_{j,k}^{n+1}}{2}
\frac{\partial \left[\left(u_i^{n}+u_{i,k}^{*,n+1}\right)/2\right]}{\partial x_j}
+
\frac{1}{\rho}\frac{\partial p_{k}}{\partial x_i}
-
\frac{\nu}{2}
\frac{\partial^2 \left(u_i^{n}+u_{i,k}^{*,n+1}\right)}
{\partial x_j \partial x_j}
-
\frac{1}{\rho}f_i.
\label{eq:momen_CN_corr}
\end{equation}

Different from the Poisson equation, the formulation of the linearized momentum equation associated with Eqs.~\eqref{eq:mom_iteration_Euler_k1} and~\eqref{eq:momen_CN_k1} changes across stages due to the pressure correction.
The pressure correction introduces additional low-frequency components to the corrected residual.
To capture both the low-frequency residual components and the high-frequency modes that remain unresolved in preceding stages, RFF embedding network structures are employed for the momentum equations at each stage.
As in the Poisson equation, the standard deviation of the RFF embedding in the first stage, $\sigma_{M_i}^{(1)}$, is set to $2.0$.
For subsequent stages, it is determined from the dominant frequency of $\hat{u}_{i,k+1}$, denoted as $f_{d,M_i}^{(k)}$, according to
\begin{equation}
\sigma_{M_i}^{(k+1)}=\frac{2\pi f_{d,M_i}^{(k)}}{3}.
\label{eq:sigma_momen}
\end{equation}
Eq.~\eqref{eq:sigma_momen} ensures that frequencies up to $f_{d,M_i}^{(k)}$ are effectively covered by the random Gaussian sampling in the RFF embedding, based on the $3\sigma$ principle.
The dominant frequency $f_{d,M_i}^{(k)}$ is evaluated from the corrected residual $\mathcal{R}_{M_i,\mathrm{corr}}^{(k)}$ using the same procedure as described in Section~\ref{sec:mpinn-Poisson}.

In Eqs.~\eqref{eq:mom_iteration_Euler_k1} and~\eqref{eq:momen_CN_k1}, the output of the newly introduced neural network, $\hat{u}_{i,k+1}$, appears in the first term, which comprises the zeroth-, first-, and second-order spatial derivative terms.
By equating the leading-order term in the first term with the magnitude of the second term in Eqs.~\eqref{eq:mom_iteration_Euler_k1} and~\eqref{eq:momen_CN_k1}, the magnitude prefactor $\epsilon_{M_i}^{(k+1)}$ in the multi-stage neural network architecture can be estimated as,
\begin{equation}
\epsilon_{M_i}^{(k+1)}
=\frac{\mathrm{RMS}\left(\mathcal{R}_{M_i,\mathrm{corr}}^{(k)}\right)
}{
\left(2\pi {f_{d,M_i}^{(k)}}\right)^n
},
\label{eq:eps_momen}
\end{equation}
where the RMS value of the corrected residual $\mathcal{R}_{M_i,\mathrm{corr}}^{(k)}$ is used to represent its magnitude and $n$ is the order of derivative in the leading order term of the first terms.

In the original multi-stage method~\cite{wang2024multi}, the magnitude prefactor is estimated solely from the equation residual, which is valid when the boundary loss is much smaller than the equation loss.
However, this condition does not hold for the momentum equation within the SIM-PINN framework.
Consequently, estimating $\epsilon_{M_i}^{(k+1)}$ solely from the equation residual using Eq.~\eqref{eq:eps_momen} would underestimate its value, preventing the boundary loss from decreasing effectively in later stages.
To address this issue, the magnitude prefactor for the momentum equation is further constrained by the average boundary residual magnitude, ensuring that boundary conditions are effectively enforced throughout all training stages.

\section{Numerical results}
\label{sec:results}
The performance of the present SIM-PINN is evaluated through the simulations of the steady lid-driven cavity flow and the unsteady Kolmogorov flow.
In the lid-driven cavity flow simulation, the performance of the multi-stage neural networks for the Poisson and momentum equations is assessed separately, before the overall performance of the fully coupled SIM-PINN framework is evaluated.
In the solution of the coupled steady and unsteady NS equations, the performance of SIM-PINN is compared with a baseline monolithic PINN and CFD reference solutions.
As presented in~\ref{app:PINN}, the baseline monolithic PINN is formulated within a similar time-marching framework as SIM-PINN.
It also employs the same RFF-embedded neural network architecture and optimization settings as those used for the momentum equation in the first stage of SIM-PINN.
The difference is that it optimizes the velocity and pressure fields simultaneously by minimizing the momentum and continuity residuals.
This setting ensures a well-configured and competitive baseline while isolating the contribution of the proposed SIM-PINN framework in the comparisons.
The detailed network architecture and training configurations for SIM-PINN and the baseline PINN are provided in~\ref{app:network_parameters}, with case-specific parameters described in their respective sections.
All test cases are non-dimensionalized and variables are dimensionless.

\subsection{Steady lid-driven cavity flows}
\label{sec:steady flows}
The lid-driven cavity flow is one of the most classical benchmark cases in CFD. 
Despite its simple geometry and boundary conditions, the lid-driven cavity flow exhibits rich flow structures, including a primary vortex, corner recirculation zones, and shearing flows near walls.
The computational domain is a 2-dimensional unit square cavity, where the top boundary moves tangentially while the other three walls remain stationary. 
To avoid the velocity discontinuity at the two upper corners where the moving lid meets the stationary side walls, the velocity on the top boundary is set as,
\begin{equation}
u_1(x,1)=u_{\mathrm{lid}}(x), \qquad u_2(x,1)=0,
\end{equation}
where
\begin{equation}
u_{\mathrm{lid}}(x)=
\begin{cases}
2\sqrt{\dfrac{x}{0.1}}-\dfrac{x}{0.1}, & 0\le x\le 0.1,\\[6pt]
1, & 0.1<x<0.9,\\[6pt]
2\sqrt{\dfrac{1-x}{0.1}}-\dfrac{1-x}{0.1}, & 0.9\le x\le 1.
\end{cases}
\label{eq:u_lid}
\end{equation}
This treatment maintains the driving effect across most of the lid while ensuring a smooth transition near the corner regions.
This reduces the negative impact of velocity singularities on the network training.
In the present work, two different Reynolds numbers\,($Re$) are considered, where $Re=1/\nu$. 
The flow at $Re=100$ is used to demonstrate the effectiveness of the multi-stage neural networks for the Poisson and momentum equations.
A more challenging case at $Re=2000$ is then used to evaluate the overall performance of SIM-PINN.

\subsubsection{Performance of segregated multi-stage PINN}
\label{sec:performance_poisson_momen}

We evaluate the multi-stage neural network performance on solving the segregated Poisson and momentum equations using lid-driven cavity flow at $Re=100$.
The initial condition is taken from the intermediate solution after 10 time steps of SIM-PINN with $\Delta t = 0.25$ and first order backward Euler time integration.
Each solver is evaluated independently: the Poisson solver updates pressure with velocity frozen, and the momentum solver updates velocity with pressure frozen. 
This decoupled setup isolates the performance characteristics of each solver component before their integration in SIM-PINN.

In each test, the performance of the multi-stage neural network, referred to as M-PINN in this subsection, is compared with that of the conventional PINN. 
For M-PINN, as described in Section~\ref{sec:mpinn}, at each stage, a new network with frequency-dependent scale factor and magnitude prefactor extracted from the previous stage is combined with previously trained networks through weighted summation to form the updated solution.
The baseline PINN uses the same network architecture but trains a single network without frequency-based adaptation. 
For fair comparison, both methods employ the same training schedule divided into three stages: 
at each stage, the Adam optimizer is used for 5000 iterations\,(epochs), followed by 5 iterations\,(epochs) of the L-BFGS optimizer.

\begin{figure} 
    \centering
    \includegraphics[width=0.60\textwidth]{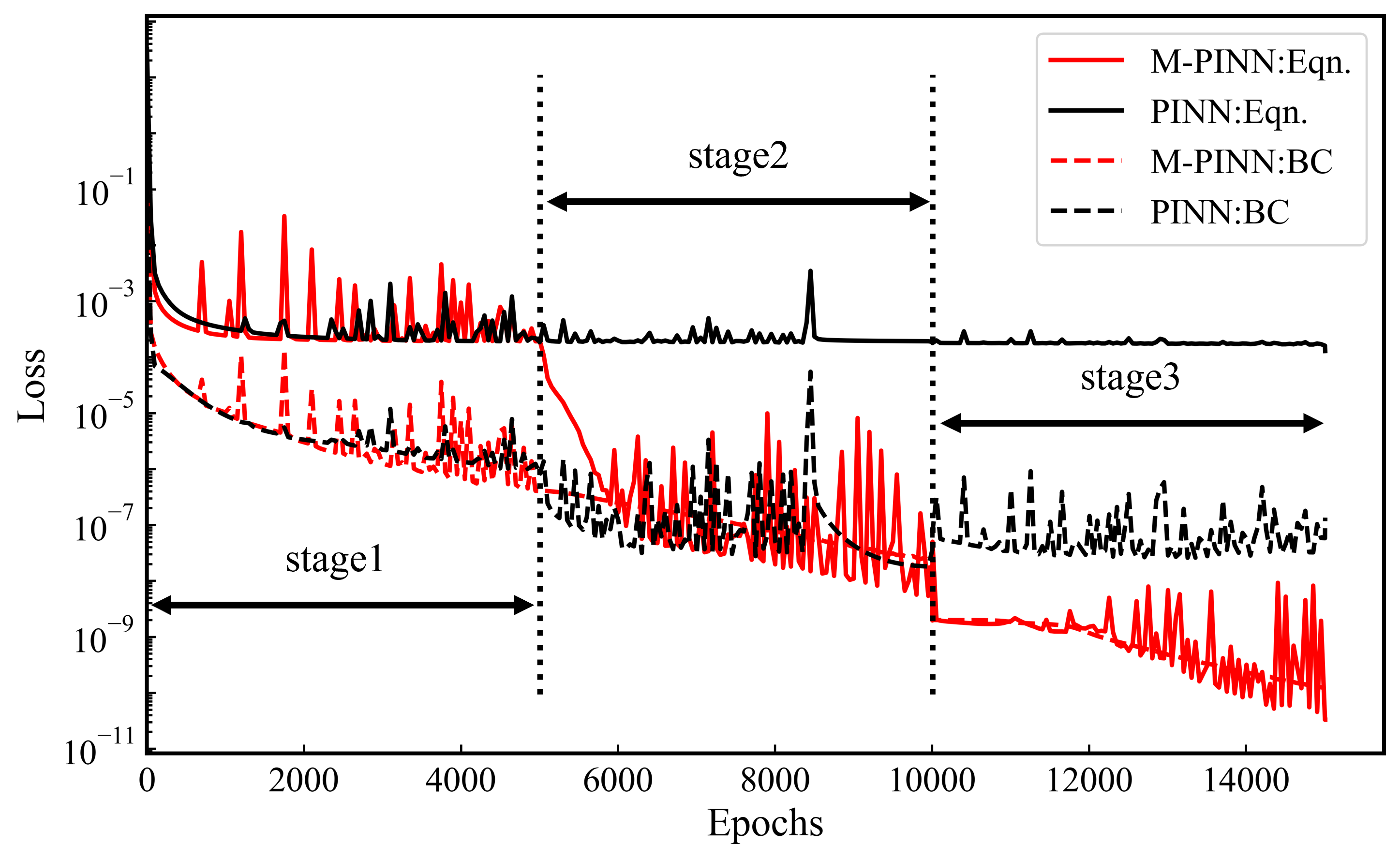}
    \caption{Comparison of losses between M-PINN and PINN during three-stage training of the Poisson equation.}
    \label{fig:poisson_loss}
\end{figure}

Figure~\ref{fig:poisson_loss} compares the loss evolution over training epochs between M-PINN and the conventional PINN for the Poisson equation.
For both methods, the boundary loss decreases rapidly, indicating that the boundary condition is not the limiting factor in solving the Poisson equation.
The primary difference lies in the evolution of the equation loss.
During the first stage, M-PINN exhibits performance similar to that of the conventional PINN, which is expected given the identical network structure.
In higher stages, M-PINN demonstrates a clear stage-wise reduction in the loss following each stage transition, with the loss continuing to decrease over epochs within each stage.
In contrast, the equation loss in the conventional PINN reaches a plateau from the second stage.
By the end of the third stage, the equation loss in M-PINN is approximately  $6$ orders of magnitude smaller than that in the conventional PINN, while the boundary loss in M-PINN is approximately $3$-orders of magnitude smaller.
This significant reduction confirms the effectiveness of the multi-stage neural network in solving the Poisson equation, consistent with Wang and Lai~\cite{wang2024multi}.

\begin{figure}  
    \centering
    \begin{subfigure}  {0.98\textwidth}
        \centering
        \includegraphics[width=0.31\textwidth]{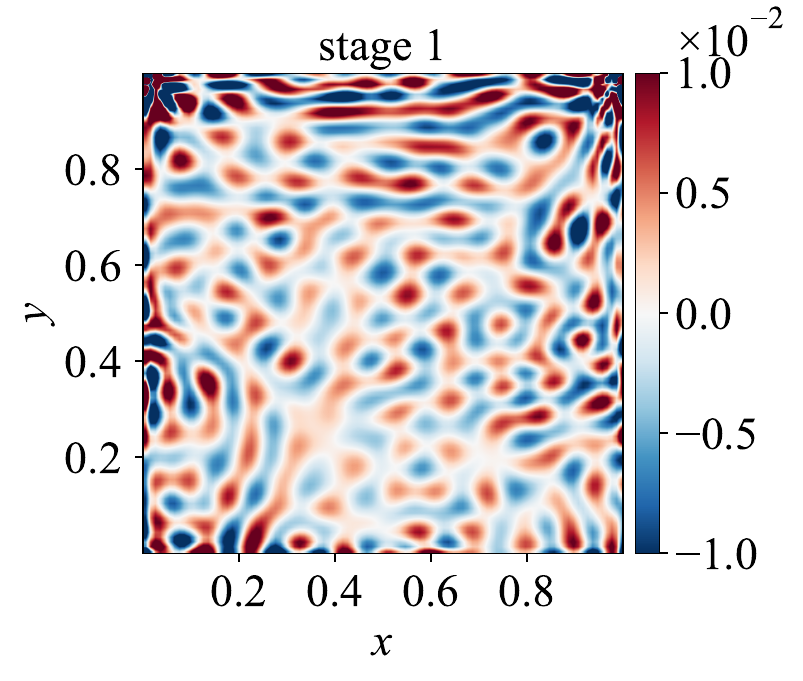}\hfill
        \includegraphics[width=0.3100\textwidth]{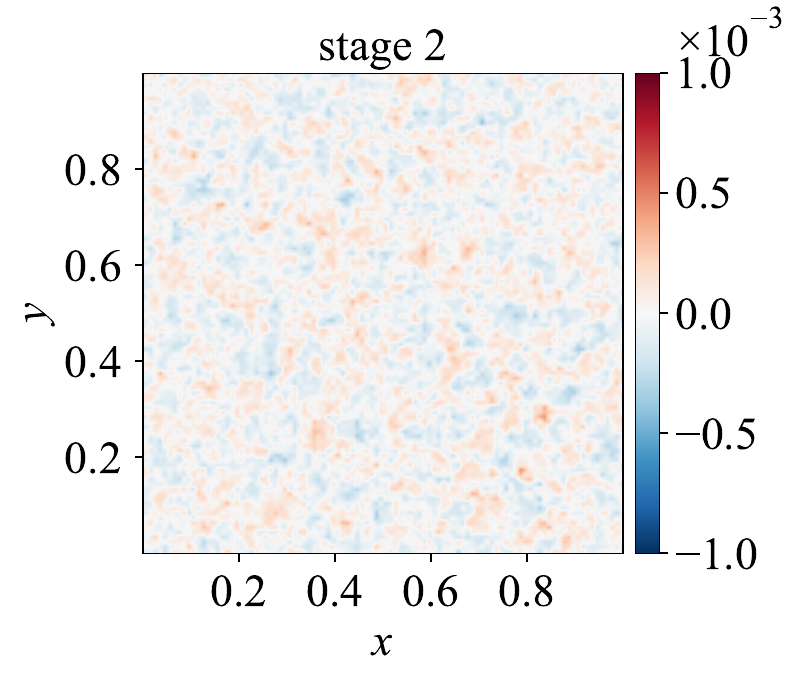}\hfill
        \includegraphics[width=0.31\textwidth]{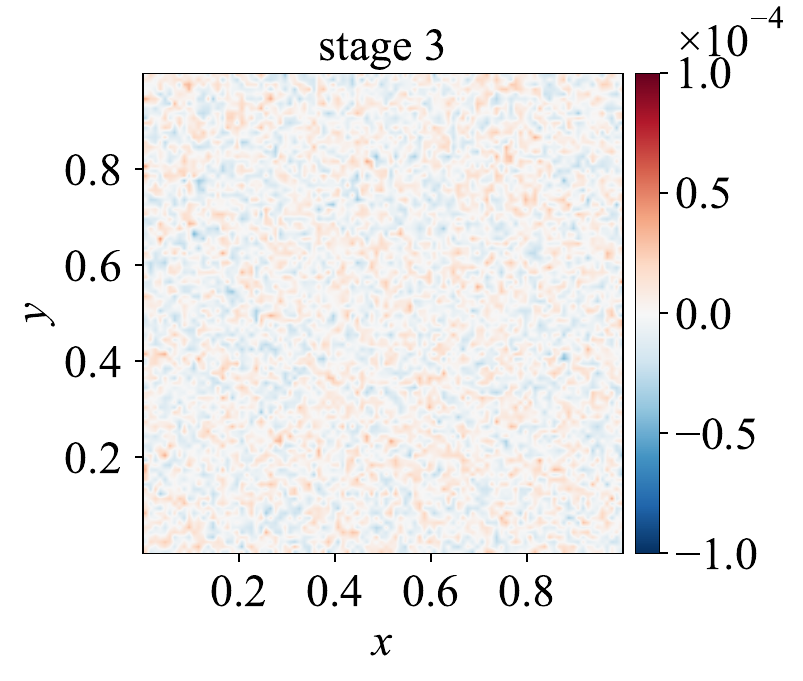}
        \caption{M-PINN}
    \end{subfigure}
    \vspace{0.4cm}
    \begin{subfigure}  {0.98\textwidth}
        \centering
        \includegraphics[width=0.31\textwidth]{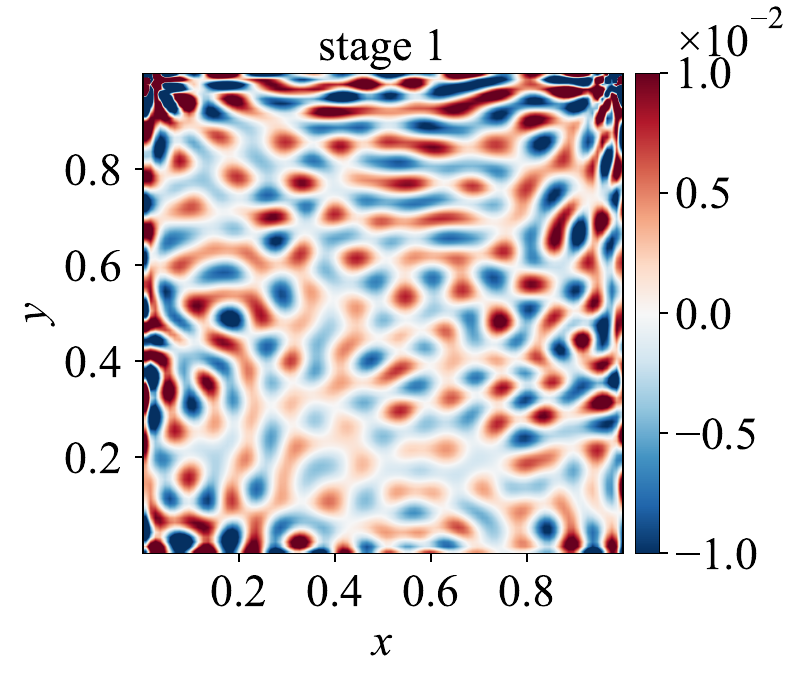}\hfill
        \includegraphics[width=0.31\textwidth]{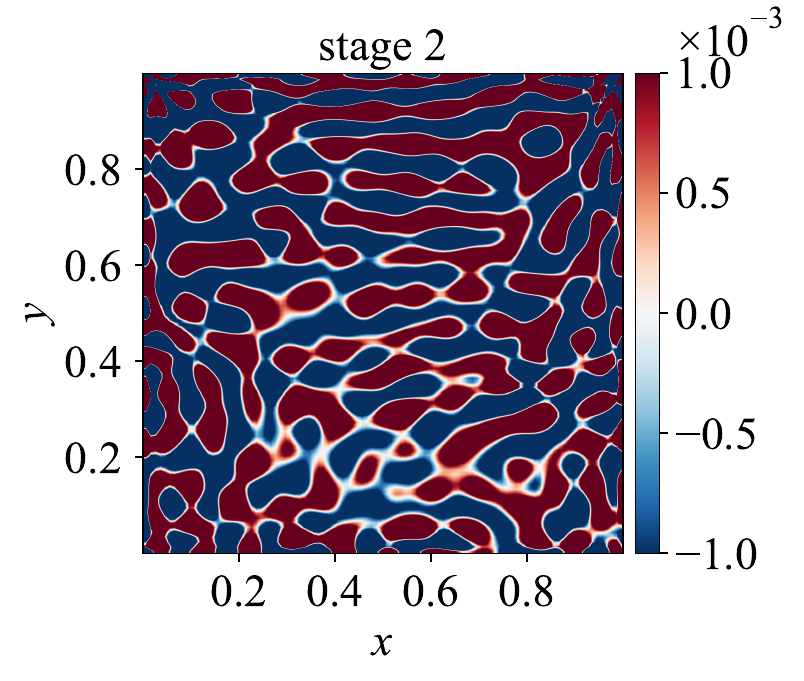}\hfill
        \includegraphics[width=0.31\textwidth]{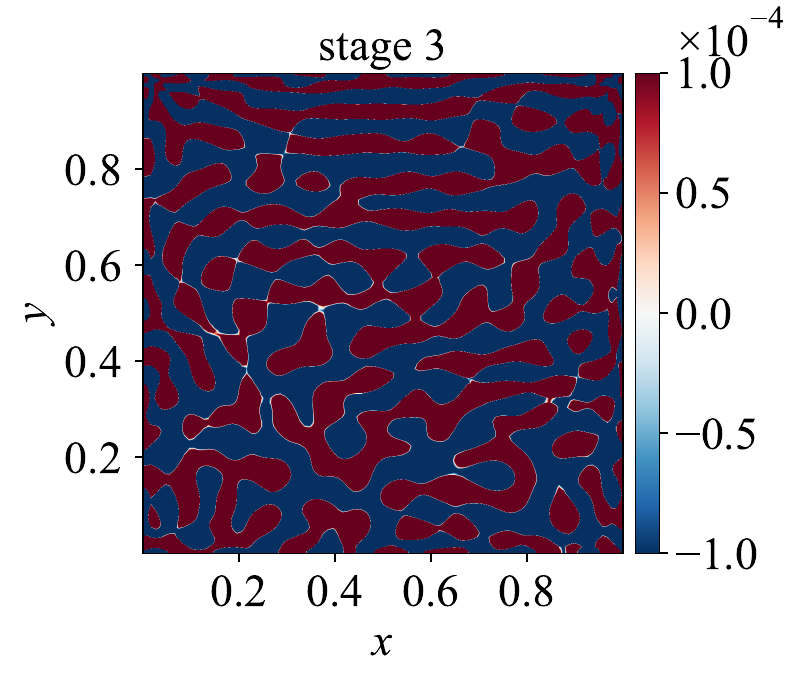}\hfill        
        \caption{PINN}
    \end{subfigure}
    \caption{Comparison of equation residual contours between M-PINN and PINN over three training stages for the Poisson equation.}
    \label{fig:poisson_residual}
\end{figure}

Figure~\ref{fig:poisson_residual} compares the contours of the Poisson equation residual between M-PINN and the conventional PINN across the three training stages. 
At the end of the first stage, the spatial distribution of the residual in M-PINN is similar to that of PINN.
As the training stage proceeds, it can be observed in M-PINN that, in addition to the reduction of the residual magnitude, the dominant frequency of the residual continuously increases, as the errors characterized by lower frequencies in the preceding stage has been effectively eliminated.
In contrast, without the frequency-based stage-wise training, the dominant frequency remains largely unchanged in the conventional PINN due to the spectral bias.

\begin{figure} 
    \centering
    \includegraphics[width=0.60\textwidth]{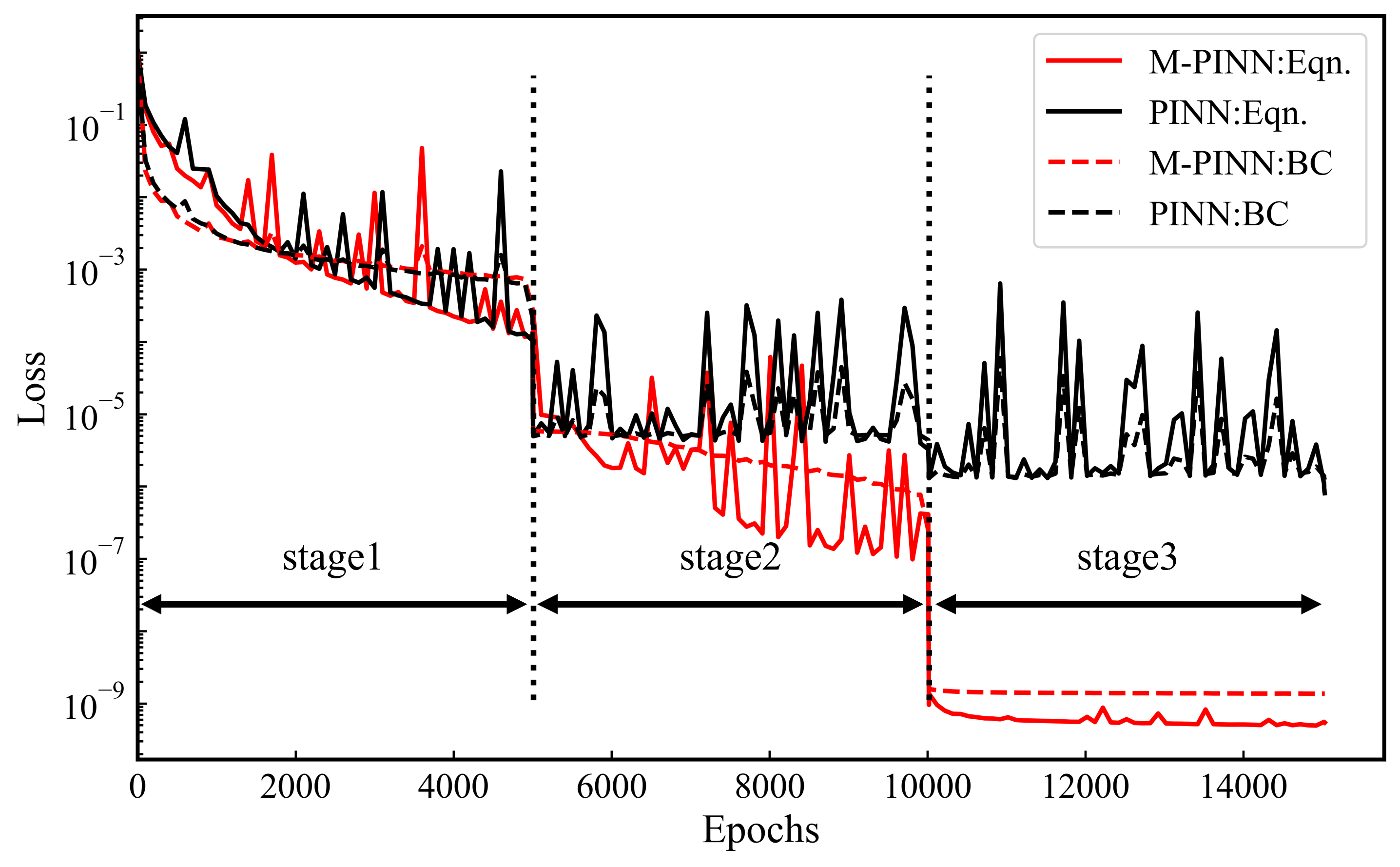}
    \caption{Comparison of losses between M-PINN and PINN during three-stage training of the momentum equation in the $x$-direction.}
    \label{fig:momentum_loss}
\end{figure}

\begin{figure} 
    \centering

    \begin{subfigure}  {0.98\textwidth}
        \centering
        \includegraphics[width=0.31\textwidth]{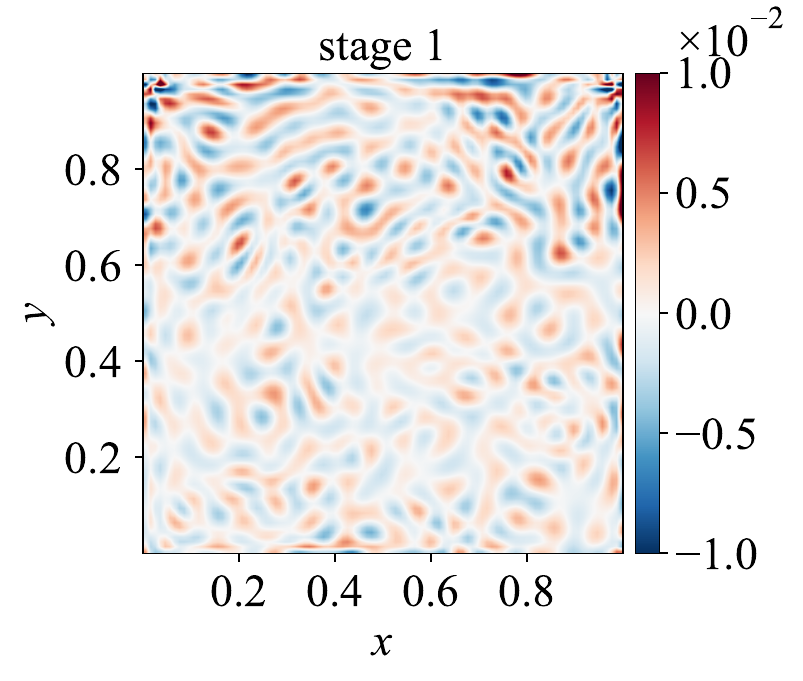}\hfill
        \includegraphics[width=0.31\textwidth]{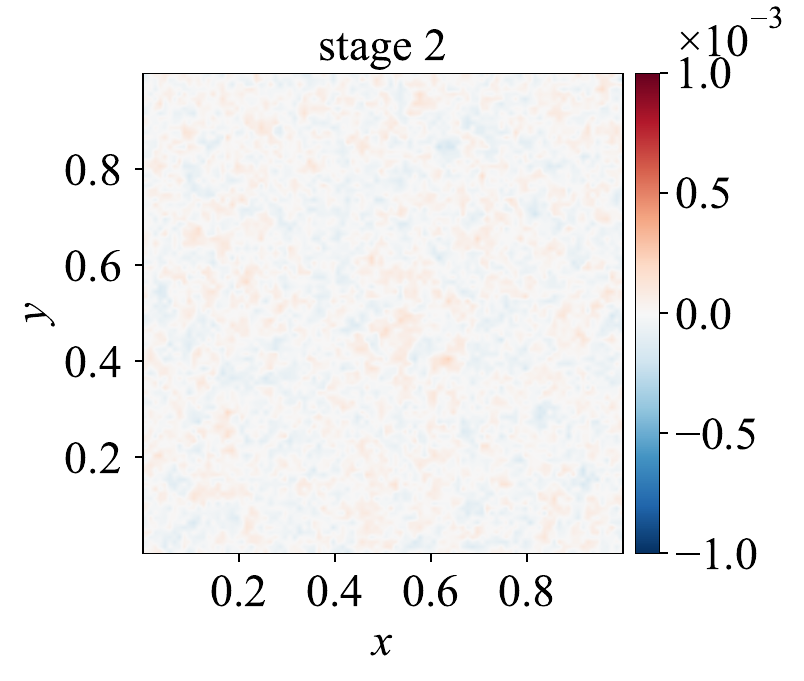}\hfill
        \includegraphics[width=0.31\textwidth]{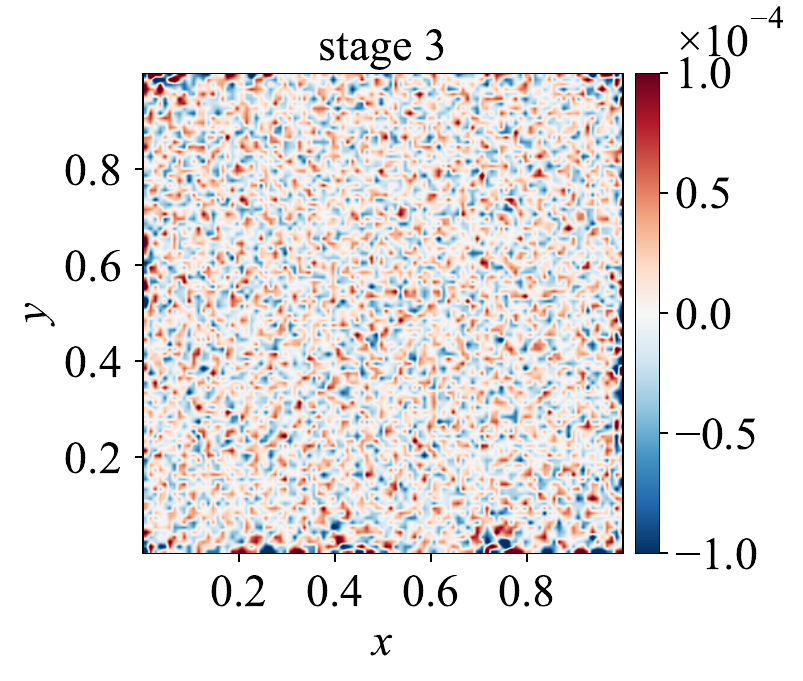}
        \caption{M-PINN}
    \end{subfigure}
    \vspace{0.4cm}
    \begin{subfigure}  {0.98\textwidth}
        \centering
        \includegraphics[width=0.31\textwidth]{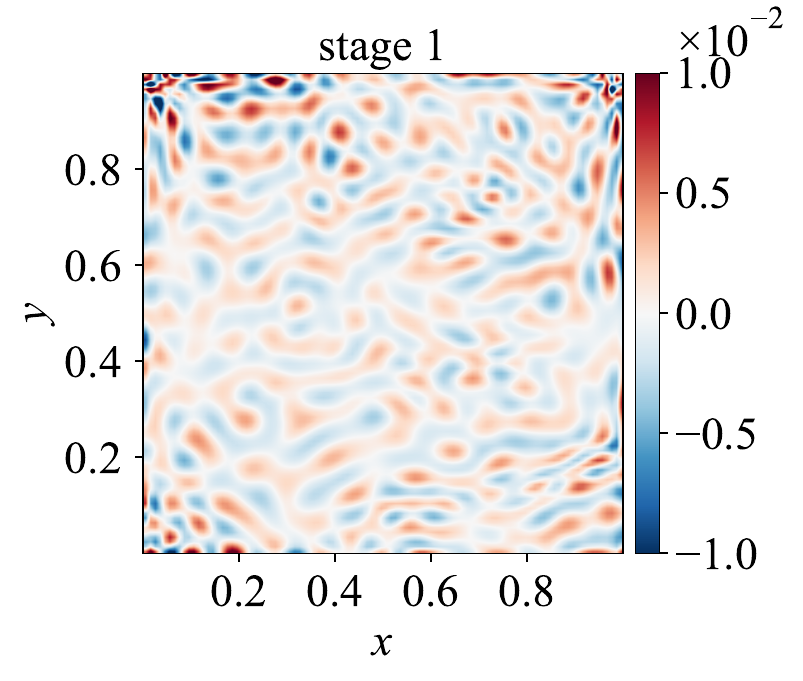}\hfill
        \includegraphics[width=0.31\textwidth]{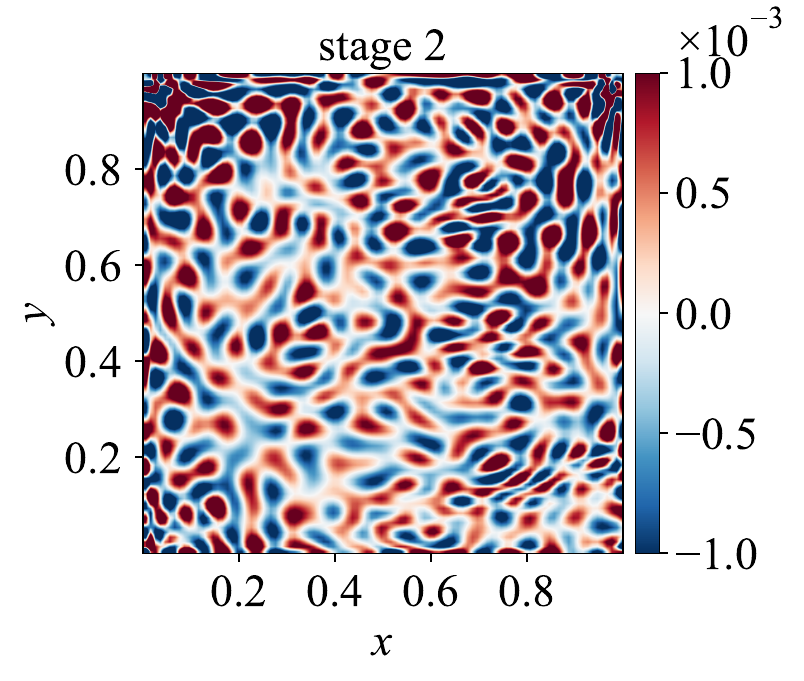}\hfill
        \includegraphics[width=0.31\textwidth]{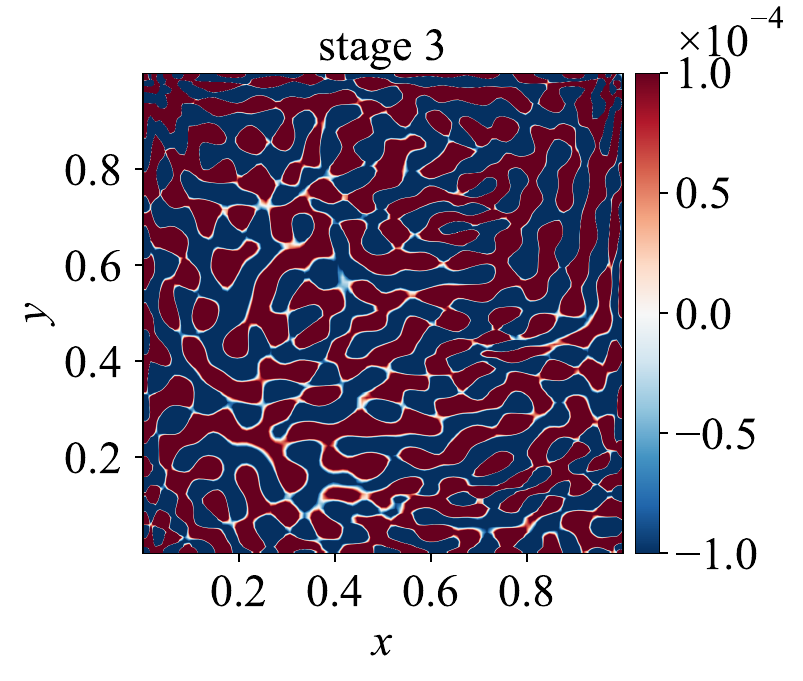}
        \caption{PINN}
    \end{subfigure}
    \caption{Comparison of equation residual contours between M-PINN and PINN over three training stages for the momentum equation in the $x$-direction.}    
    \label{fig:momentum_residual}
\end{figure}

Figures~\ref{fig:momentum_loss} and~\ref{fig:momentum_residual} compare the loss evolution and the spatial distribution of the equation residuals, respectively, for the momentum equation between M-PINN and the conventional PINN.
Here, results for the momentum equation in the $x$-direction are presented.
And the results for the $y$-direction momentum equation exhibit similar behaviour.
While the momentum equation is more complex than the Poisson equation, the overall qualitative difference between M-PINN and the conventional PINN remains similar to that observed for the Poisson equation.
By the end of the third stage, both the equation loss and the boundary loss in M-PINN are more than $3$ orders of magnitude smaller than those in the conventional PINN.
In addition, the dominant frequency of the residual in M-PINN continuously increases, whereas that in the conventional PINN remains stagnant.

It is observed in Fig.~\ref{fig:momentum_loss} that, in the conventional PINN, the boundary loss remains on the same order of magnitude as the equation loss, even after sufficient training epochs.
This observation justifies the design choice of bounding the magnitude prefactor $\epsilon_{M_i}^{(k+1)}$ by the average boundary residual magnitude, as described in Section~\ref{sec:mpinn-momen}, thereby ensuring effective boundary condition enforcement.

\subsubsection{Performance of SIM-PINN}
\label{sec:steady_simu_SIM}

To assess the performance of the proposed SIM-PINN for solving the coupled steady-state Navier-Stokes equations at relatively high Reynolds numbers, the lid-driven cavity flow at $Re=2000$ is considered. 
The network architecture and training optimization parameters are detailed in~\ref{app:network_parameters}.
The number of sub-iterations\,(stages) in SIM-PINN is set to $2$, and the pseudo-time step size is set to $\Delta t=0.25$.
A total of $300$ pseudo-time steps are performed to reach the steady state.
The results obtained by SIM-PINN are compared with those from the baseline monolithic PINN and a reference solution computed using OpenFOAM~\cite{weller1998tensorial,greenshields2023openfoam11}.
The baseline PINN, whose formulation is detailed in~\ref{app:PINN}, employs the same RFF-embedded neural network architecture and optimization settings as those used for the momentum equation in the first stage of SIM-PINN.
Unlike SIM-PINN, it optimizes the velocity and pressure fields simultaneously without the segregated and iteratively coupled solution procedure or frequency-adapted stage-wise training.
At each pseudo-time step, the Adam optimizer is applied for $5000$ epochs, followed by $5$ epochs of the L-BFGS optimizer.
The total number of pseudo-time steps for the baseline PINN is set to $1000$, with the same time step size as SIM-PINN, to ensure sufficient convergence.
Both the monolithic PINN and SIM-PINN employ $40{,}000$ interior collocation points uniformly distributed in the computational domain.
In addition, $200$ boundary collocation points are sampled on each boundary side to enforce the boundary conditions.
For the reference CFD solution, a uniform Cartesian grid of $400\times 400$ is used, for which grid convergence has been verified.

\begin{figure} 
    \centering

    \begin{subfigure}  {0.325\textwidth}
        \centering
        \includegraphics[width=\textwidth]{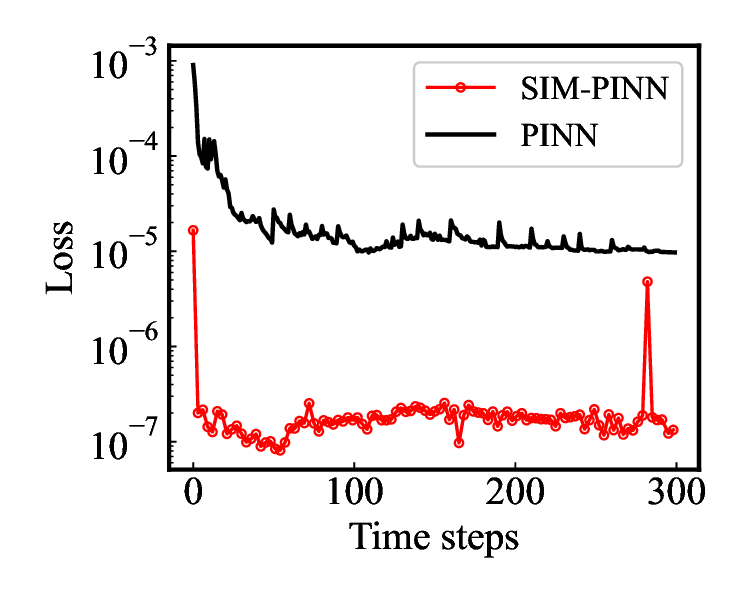}
        \caption{$x$-momentum equation}
        \label{fig:mse-u}
    \end{subfigure}
    \begin{subfigure}  {0.325\textwidth}
        \centering
        \includegraphics[width=\textwidth]{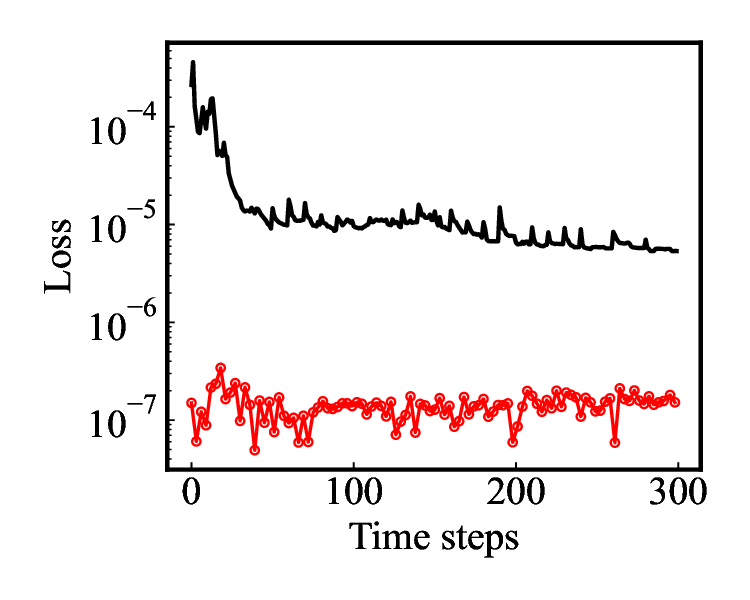}
        \caption{$y$-momentum equation}
        \label{fig:mse-v}
    \end{subfigure}
    \begin{subfigure}  {0.325\textwidth}
        \centering
        \includegraphics[width=\textwidth]{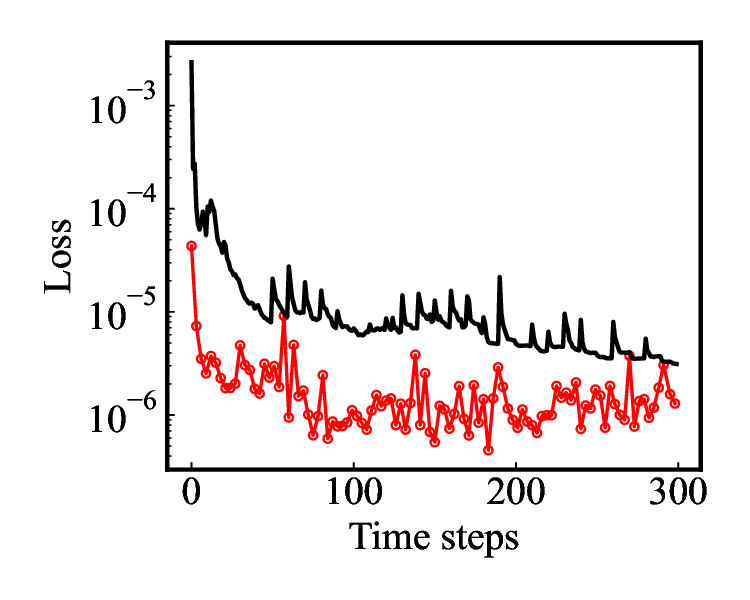}
        \caption{continuity equation}
        \label{fig:mse-div}
    \end{subfigure}
    \caption{Comparison of the loss evolution between SIM-PINN and PINN over the first 300 pseudo-time steps for the lid-driven cavity flow at $Re=2000$.}
    \label{fig:mse-history}
\end{figure}

Figure~\ref{fig:mse-history} compares the evolution of the total loss, which is the sum of the equation and boundary losses, between SIM-PINN and the conventional monolithic PINN over the first $300$ pseudo-time steps.
The loss reported for SIM-PINN at each time step is taken after the final inner sub-iteration of that step.
The results show that SIM-PINN consistently maintains lower loss levels for the $x$- and $y$-momentum equations as well as for the divergence constraint compared to the monolithic PINN, which is consistent with the results of the segregated M-PINN reported in Section~\ref{sec:performance_poisson_momen}.
In particular, the momentum-equation residuals are reduced by approximately two orders of magnitude relative to those of the monolithic PINN.

\begin{figure} 
    \centering

    \begin{subfigure}[b]{0.98\textwidth}
        \centering
        \includegraphics[width=0.31\textwidth]{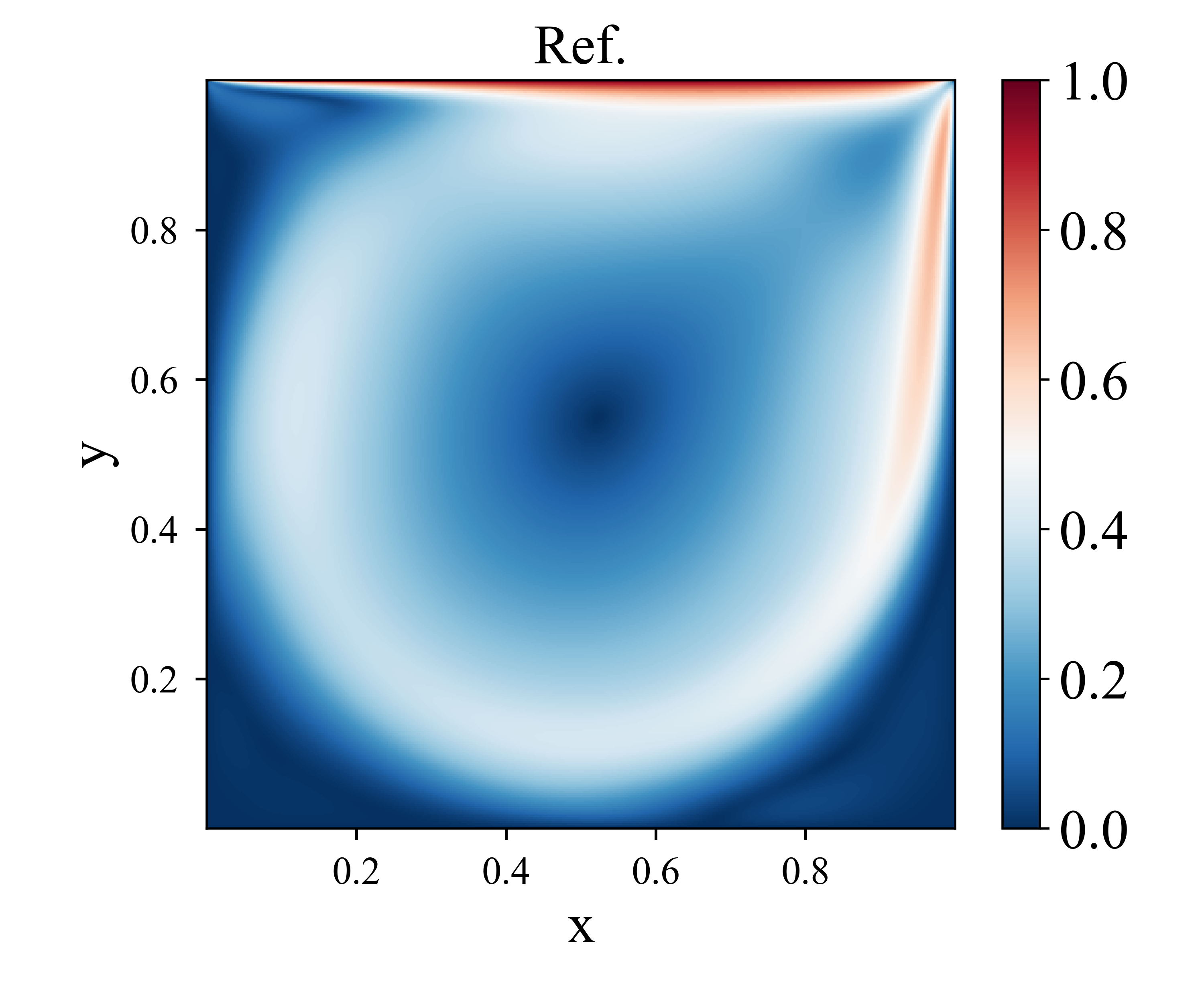}\hfill
        \includegraphics[width=0.31\textwidth]{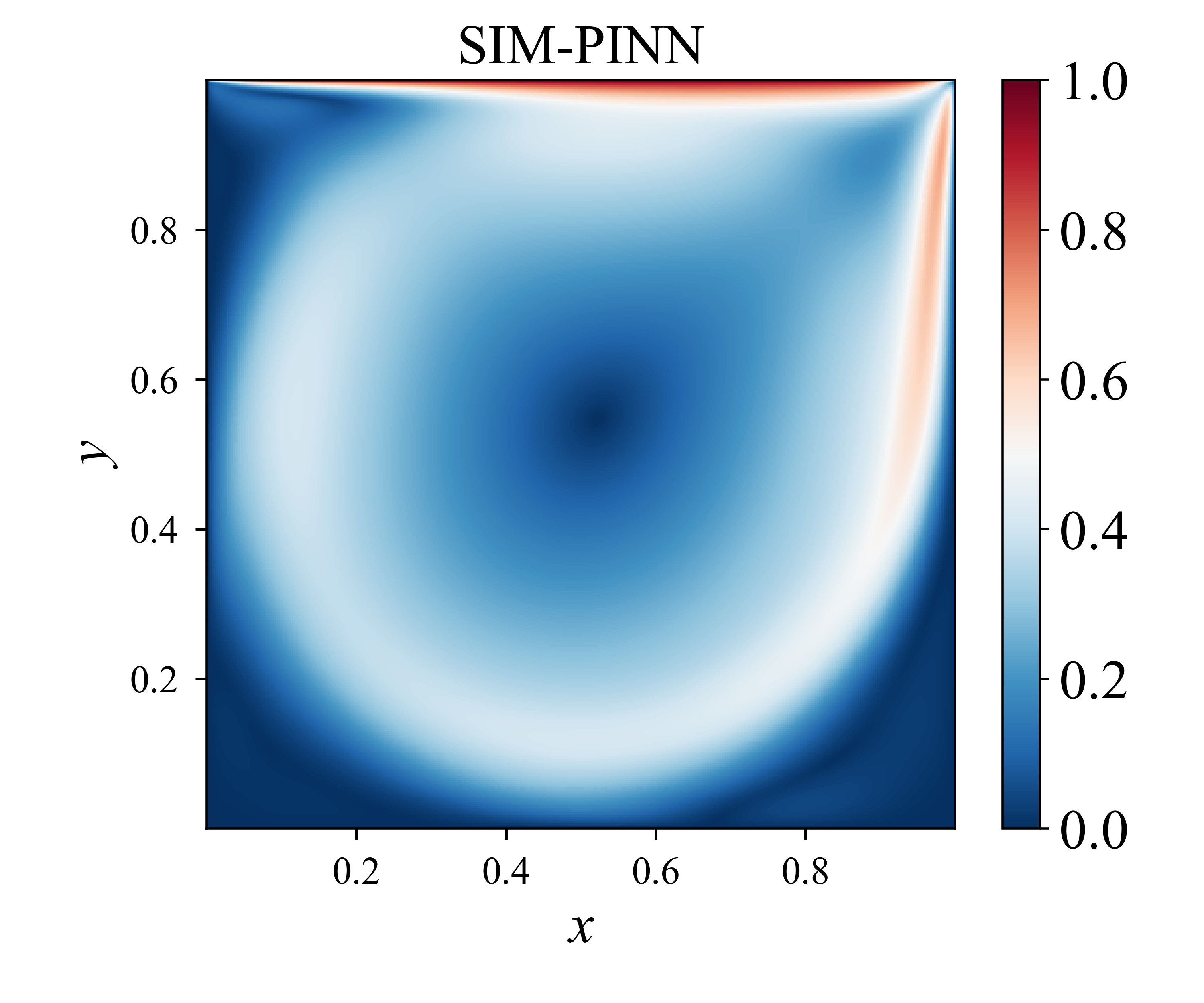}\hfill
        \includegraphics[width=0.31\textwidth]{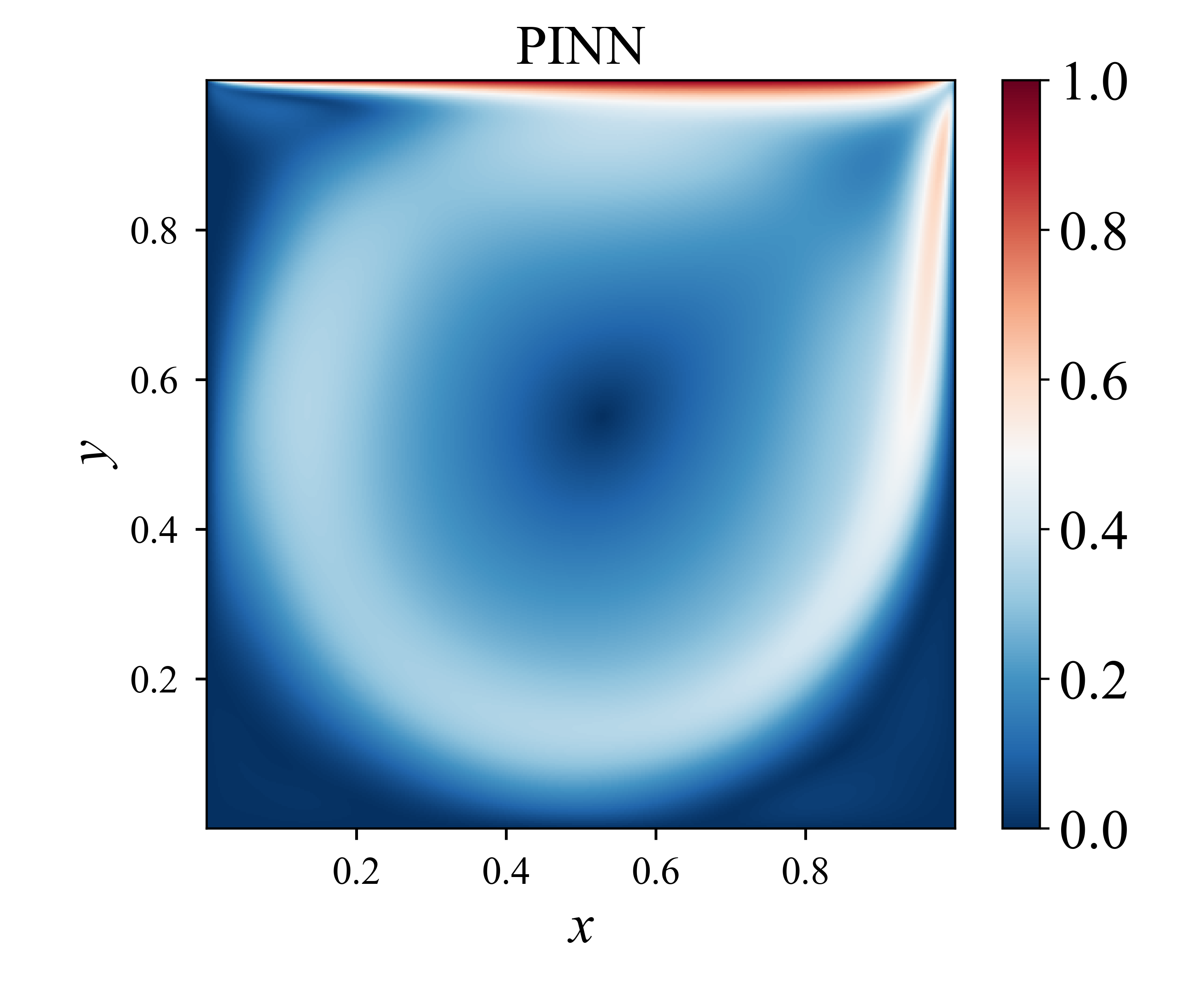}
        \caption{velocity magnitude}
        \label{fig:row-a}
    \end{subfigure}

    \vspace{0.5cm}

    \begin{subfigure}[b]{0.98\textwidth}
        \centering
        \makebox[0.31\textwidth]{}\hfill
        \includegraphics[width=0.31\textwidth]{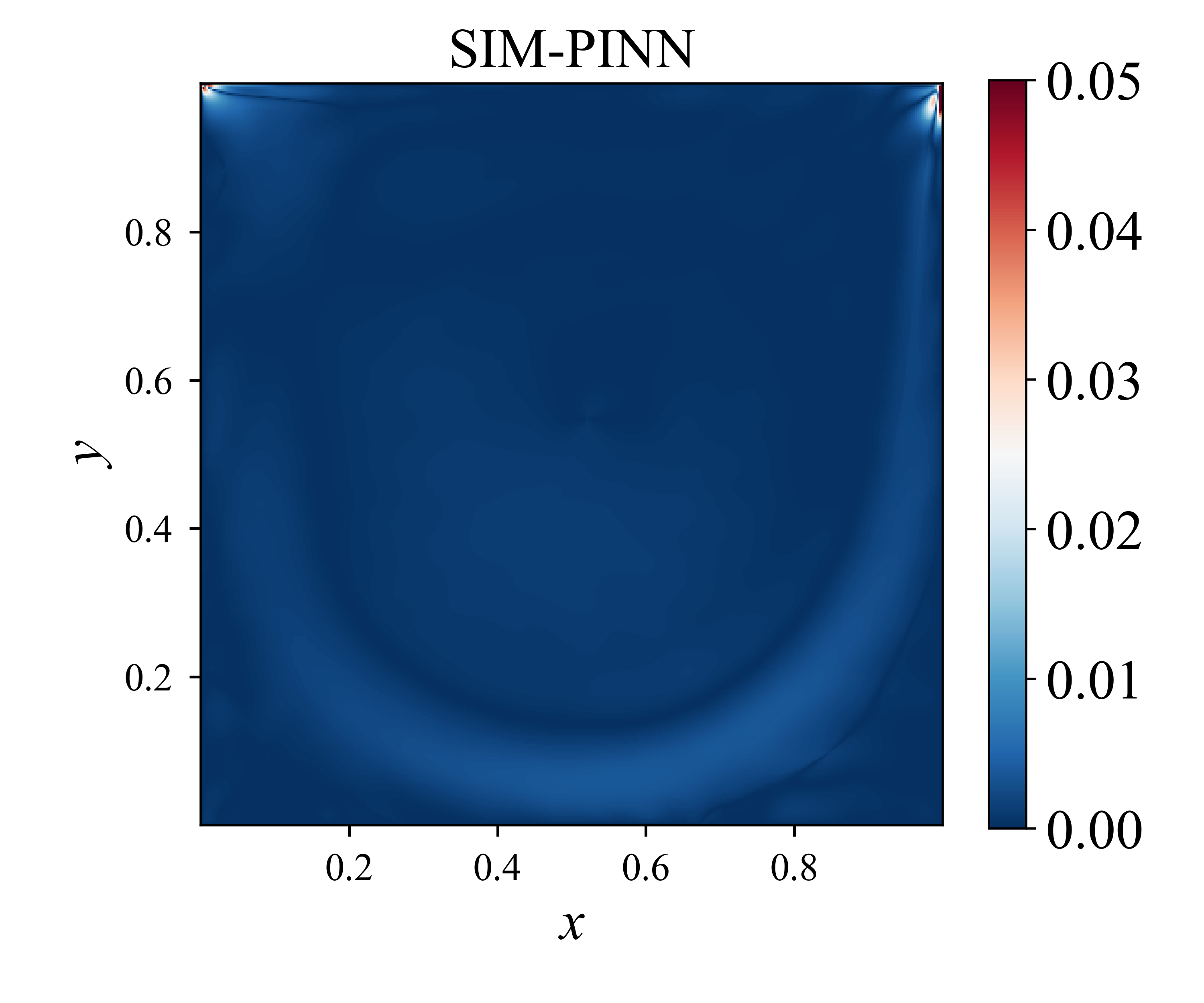}\hfill
        \includegraphics[width=0.31\textwidth]{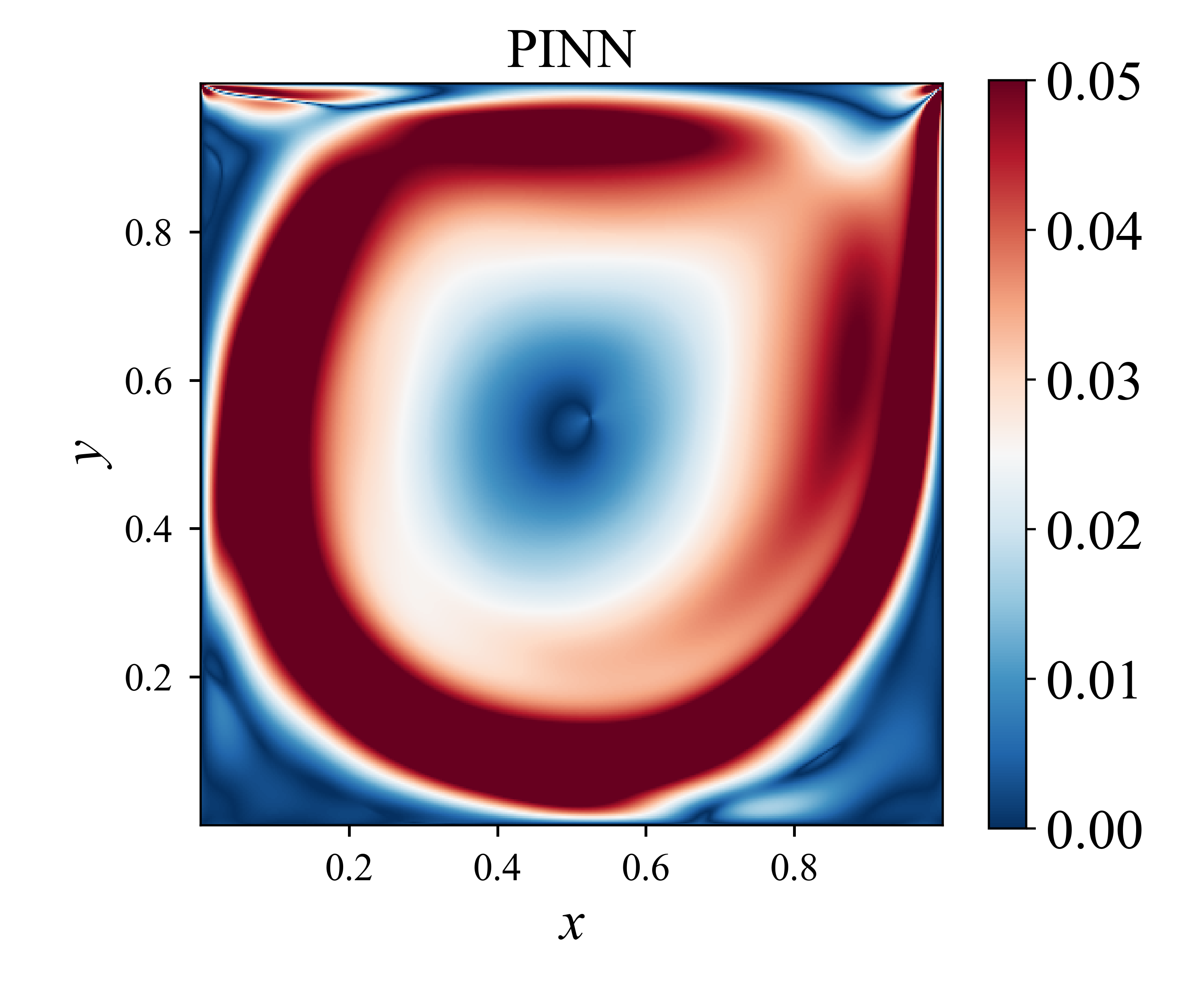}
        \caption{absolute error}
        \label{fig:row-b}
    \end{subfigure}

    \caption{Comparison of velocity magnitude\,(a) and error\,(b) contours between SIM-PINN and PINN for the lid-driven cavity flow at $Re=2000$.}
    \label{fig:cavity_erro}
\end{figure}

\begin{figure} 
    \centering

    \begin{subfigure}[b]{0.48\textwidth}
        \centering
        \includegraphics[width=\textwidth]{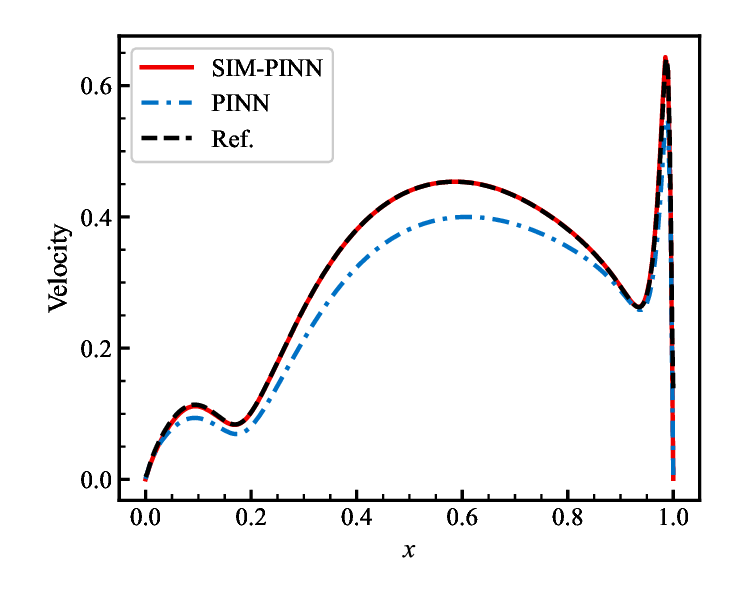}
        \caption{$y=0.95$}
        \label{fig:sub-095}
    \end{subfigure}
    \hfill
    \begin{subfigure}[b]{0.48\textwidth}
        \centering
        \includegraphics[width=\textwidth]{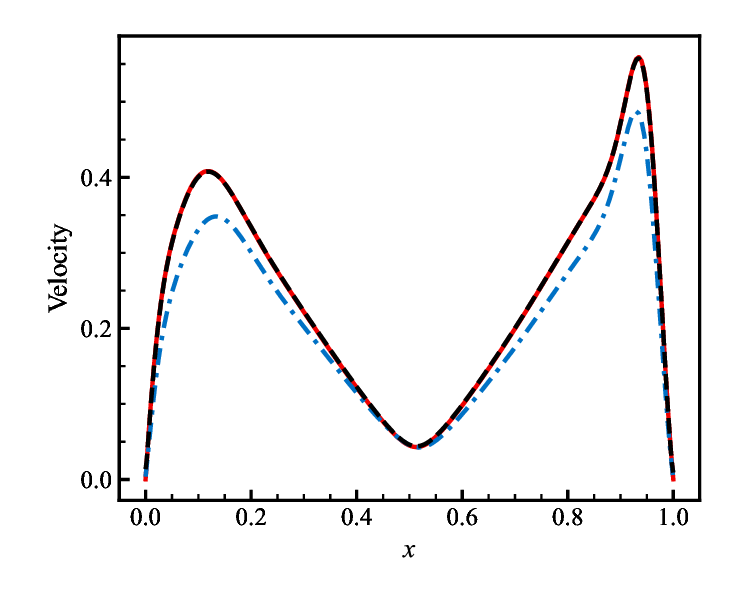}
        \caption{$y=0.5$}
        \label{fig:sub-05}
    \end{subfigure}


    \begin{subfigure}[b]{0.48\textwidth}
        \centering
        \includegraphics[width=\textwidth]{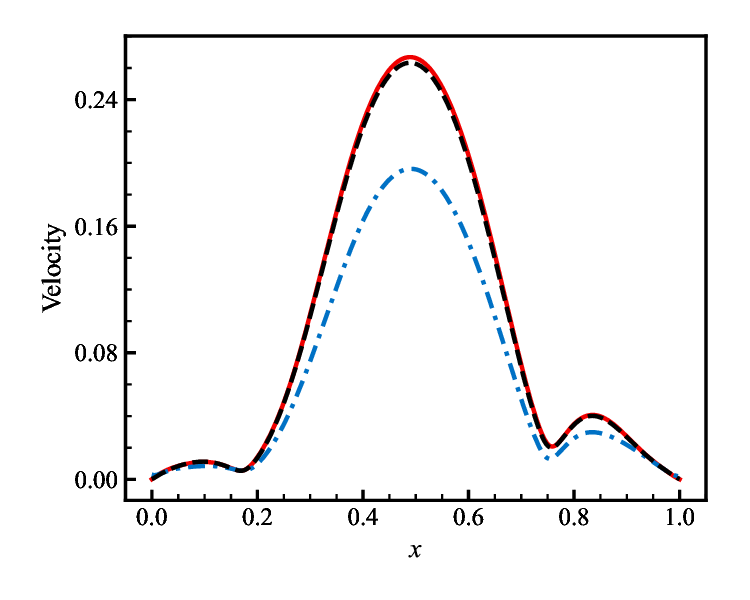}
        \caption{$y=0.05$}
        \label{fig:sub-005}
    \end{subfigure}
    \hfill
    \begin{subfigure}[b]{0.48\textwidth}
        \centering
        \includegraphics[width=\textwidth]{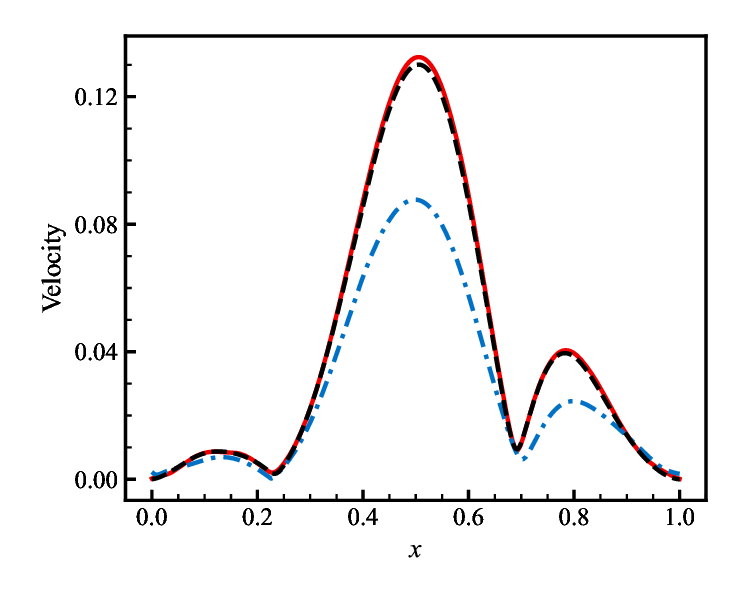}
        \caption{$y=0.02$}
        \label{fig:sub-002}
    \end{subfigure}
    \caption{Comparison of velocity profiles along different vertical locations in the lid-driven cavity flow at $Re=2000$.}
    \label{fig:Comparison of velocity}
\end{figure}

Figure~\ref{fig:cavity_erro} shows the predicted velocity magnitude and the corresponding absolute error distributions with respect to the reference solution for the lid-driven cavity flow at $Re=2000$. 
The flow exhibits a large primary vortex occupying the cavity center and multiple secondary corner vortices with low velocities in the bottom and top-left corners.
While both SIM-PINN and the monolithic PINN capture the overall flow pattern qualitatively, clear differences appear in the near-wall and corner regions.
As shown in Fig.~\ref{fig:cavity_erro}\,(b), noticeable deviations remain near the walls and corners for the results of the monolithic PINN.
In contrast, SIM-PINN produces a more uniform error distribution with much smaller magnitude.
The difference between SIM-PINN and PINN is further demonstrated in Fig.~\ref{fig:Comparison of velocity}, which shows the velocity profiles along different vertical locations.
The results of the proposed SIM-PINN are much closer to the reference solutions, particularly near the local maxima, which correspond to either shearing flows near the walls or the small-scale secondary vortex structures near the corners.
\textcolor{black}{It is noted that increasing the number of pseudo-time steps for the monolithic PINN does not yield further improvements.}
Two additional experiments with the monolithic PINN are presented in~\ref{app:cost_matched_comparison}: one with a computational cost comparable to that of SIM-PINN and the other with the same number of trainable parameters as SIM-PINN.
Although the training losses are further reduced, the velocity-field prediction accuracy remains appreciably lower than that achieved by SIM-PINN.
The improved accuracy of SIM-PINN suggests that it is more suitable for simulating incompressible flows at high Reynolds numbers.

\subsection{Unsteady Kolmogorov flow}
\label{sec:unsteady flows}
The two-dimensional Kolmogorov flow, which has been widely used to study unsteady vortical dynamics~\cite{lucas2014spatiotemporal,lucas2015recurrent} and forced turbulence~\cite{chandler2013invariant}, is employed to assess the performance of SIM-PINN in simulating unsteady incompressible flows.
This flow is defined on a periodic domain $\Omega=[0,1]^2$ and driven by a sinusoidal body force $\mathbf{f}(x,y)=\bigl(\sin(2\pi n y),\,0\bigr)$, where $n$ is the forcing wavenumber. 
The force acts in the $x$-direction with an initially dominant length scale in the $y$ direction.
In the present study, $n=4$ and $Re=640$ are considered.
Despite its simple configuration, this flow exhibits complex unsteady features including nonlinear instability and multi-scale coherent structures, making it a challenging benchmark for evaluating the capability of PINN methods~\cite{wang2025simulating}.

The network architecture and training optimization parameters for SIM-PINN are identical to those used in the steady lid-driven cavity flow case and are detailed in~\ref{app:network_parameters}.
To maintain temporal accuracy, the second-order Crank--Nicolson scheme is employed with a constant time step size of $0.025$.
A total of $40{,}000$ uniformly distributed interior collocation points are sampled in the computational domain, and periodic boundary conditions are enforced through periodic mapping.
The simulation is carried out up to a total physical time of $T=4$, equivalent to $160$ time steps.
At each time step, the number of sub-iterations is determined adaptively using a residual-based convergence criterion~\cite{li2025learning}, such that the losses of the corrected residual for the linearized momentum equations fall below $1\times10^{-5}$.
The number of stages for the Poisson equation is set to $2$.
The predictions of SIM-PINN are compared with reference CFD results obtained using a high-order finite difference CFD solver~\cite{Desjardins2008,Su2018}.
For the reference CFD solution, a uniform Cartesian grid of $200\times 200$ is employed, which has been verified to be sufficiently refined through a grid convergence study, and the CFL number is fixed at $0.5$.
The initial field in CFD is set to the same as in SIM-PINN, using a randomly generated velocity field.
The baseline monolithic PINN, employing the same RFF-embedded neural network architecture, optimization settings, and time step size as those used for the momentum equation in the first stage of SIM-PINN, fails to converge to a physically meaningful solution.
Consequently, the baseline PINN results are not presented for comparison.

\begin{figure}
    \centering

    \begin{subfigure}[b]{0.32\textwidth}
        \centering
        \includegraphics[width=\textwidth]{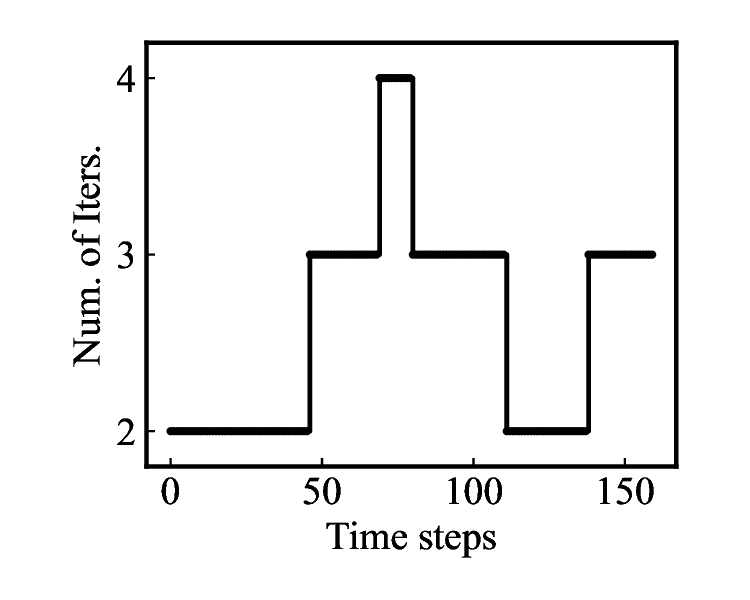}
        \caption{Number of inner iterations}
        \label{fig:kolmogorov_kinner}
    \end{subfigure}%
    \hfill
    \begin{subfigure}[b]{0.32\textwidth}
        \centering
        \includegraphics[width=\textwidth]{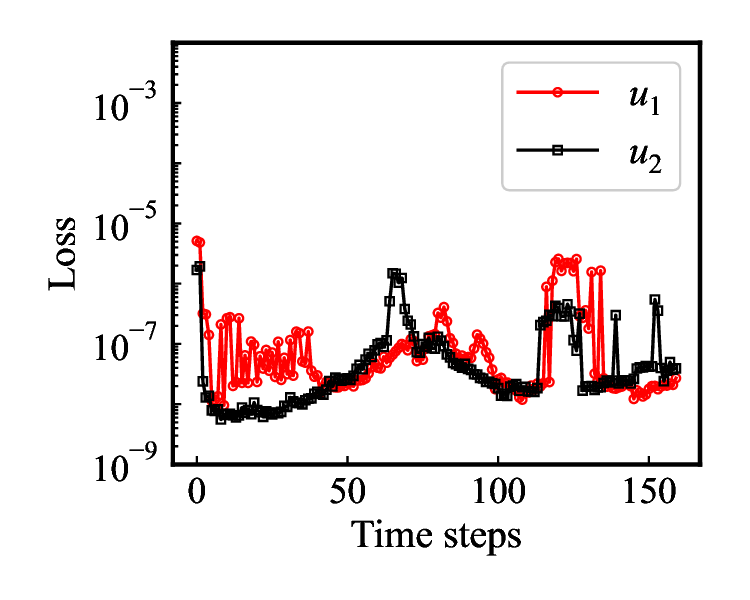}
        \caption{Momentum equation}
        \label{fig:kolmogorov_resv}
    \end{subfigure}%
    \hfill
    \begin{subfigure}[b]{0.32\textwidth}
        \centering
        \includegraphics[width=\textwidth]{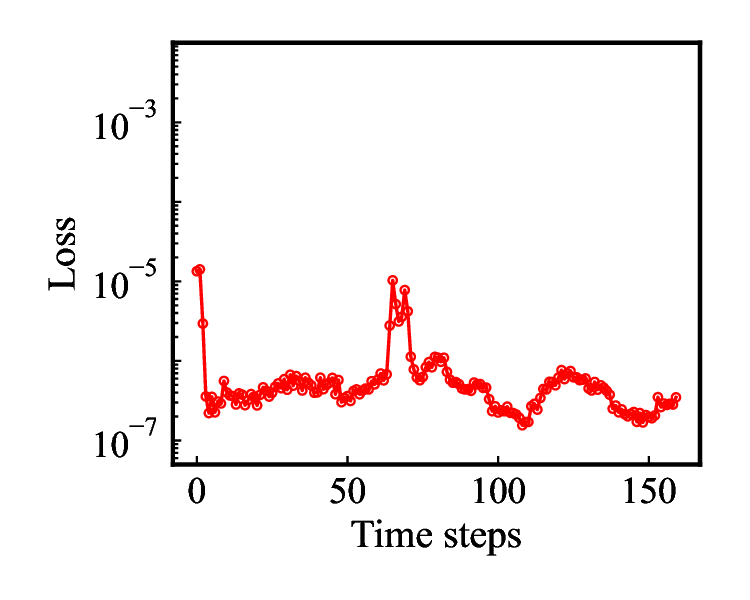}
        \caption{Divergence}
        \label{fig:kolmogorov_DIV}
    \end{subfigure}
\caption{Number of inner iterations and loss values at different time steps for the two-dimensional Kolmogorov flow with SIM-PINN.}
    \label{fig:kolmogorov_iter_res}
\end{figure}

Figure~\ref{fig:kolmogorov_iter_res} presents the required number of sub-iterations and the losses for each equation obtained with SIM-PINN at each physical time step.
For most time steps, SIM-PINN requires only 2--3 sub-iterations and the corresponding losses of the momentum equations typically remain below $10^{-6}$.
The divergence residual also remains well controlled, indicating that the incompressibility constraint is maintained throughout the simulation.

\begin{figure}
    \centering
    \includegraphics[width=0.98\textwidth]{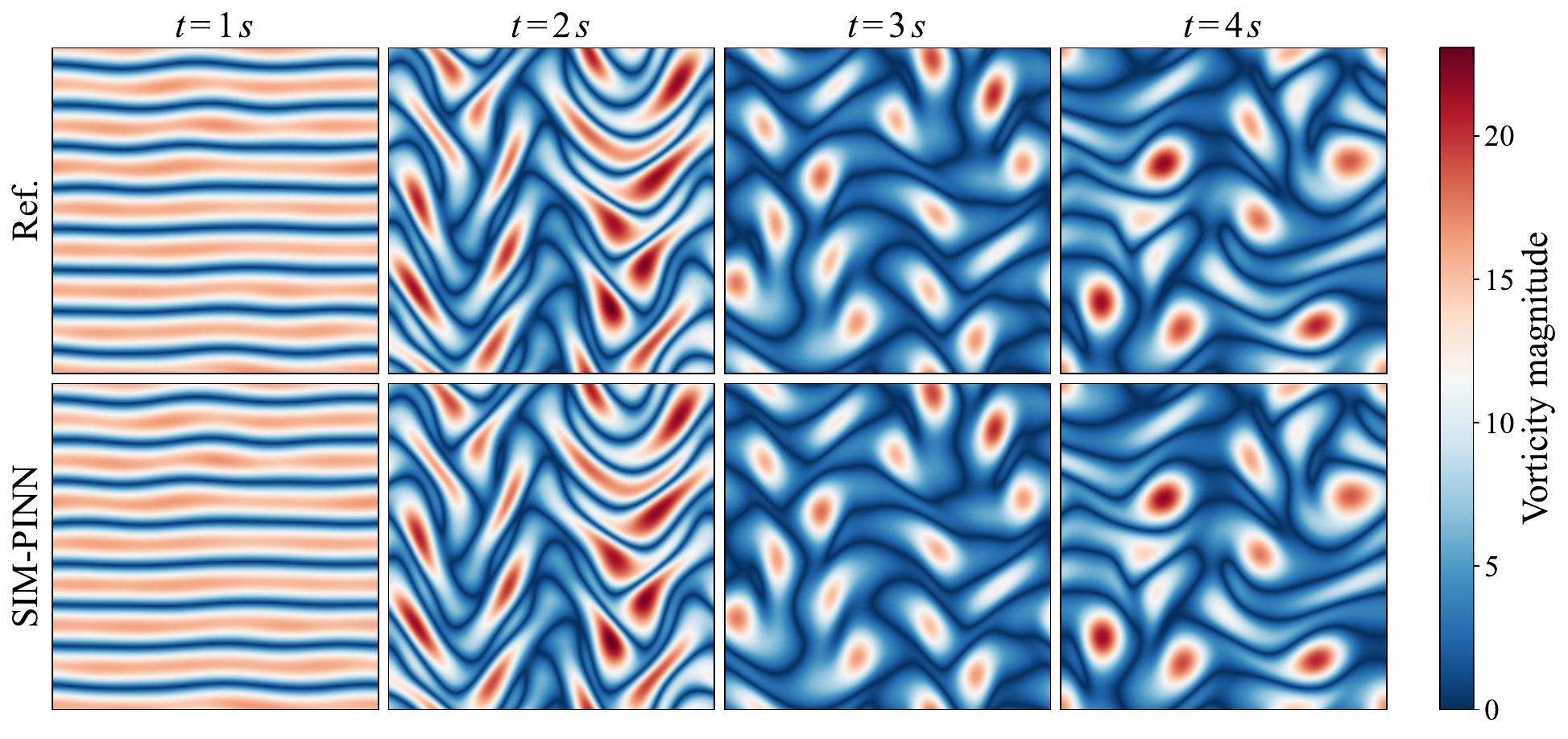}
    \caption{Vorticity snapshots of the two-dimensional Kolmogorov flow at different time instants.}
    \label{fig:kolmogorov_vorticity}
\end{figure}

Figure~\ref{fig:kolmogorov_vorticity} shows vorticity snapshots at different time instants during the flow evolution. 
At $t=1$, the flow exhibits horizontal alternating vorticity bands with weak, wave-like undulations, indicating the onset of Kelvin--Helmholtz instability. 
At $t=2$, the vorticity bands begin to break up, forming pronounced roll-up regions and locally detached vortical structures. 
At times $t=3$ and $t=4$, the flow transitions to a complex state comprising multi-scale vortical structures.
The original vorticity bands have fragmented into discrete vortex cells exhibiting irregular spatial patterns.
The flow evolution at all instances are accurately reproduced by SIM-PINN without observable phase shift or structural distortion. 

Figure~\ref{fig:kolmogorov_stats} presents the temporal evolution of enstrophy, total kinetic energy, and the relative $L_2$ error of velocity magnitude predicted by SIM-PINN.
The relative $L_2$ error is computed as the $L_2$ norm of the difference between the SIM-PINN prediction and the reference CFD solution, normalized by the $L_2$ norm of the latter.
The predicted enstrophy and kinetic energy are in close agreement with the reference solution. 
Both quantities increase during the early forcing-dominated stage and subsequently decrease as the flow develops into multi-scale vortical structures. 
The relative $L_2$ error of SIM-PINN remains low over the studied time period, without noticeable long-term drift.
The close agreement between the result of SIM-PINN and the reference solution suggests that SIM-PINN is capable of accurately simulating unsteady, complex, and multi-scale incompressible turbulent flows.

\newlength{\kolfigheight}
\setlength{\kolfigheight}{4.0cm}
\begin{figure}
    \centering

    \begin{subfigure}[b]{0.314\textwidth}
        \centering
        \includegraphics[height=\kolfigheight,keepaspectratio]{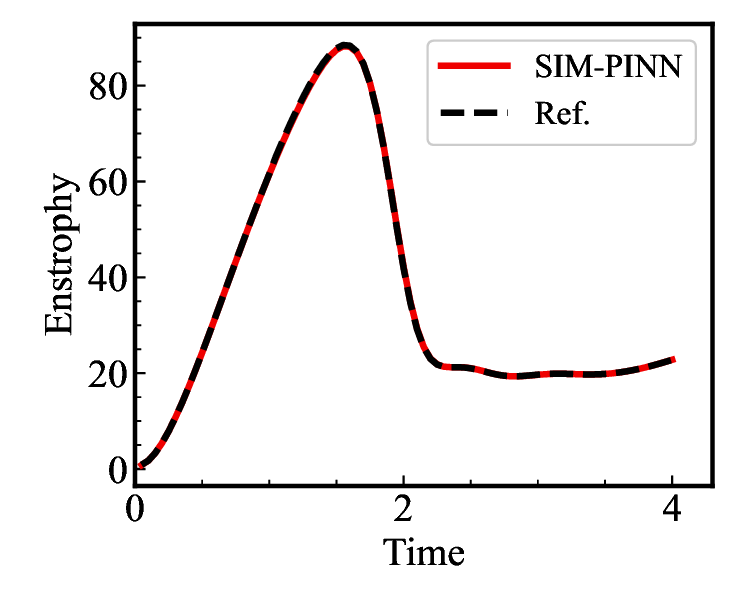}
        \caption{Enstrophy}
        \label{fig:kolmogorov_enstrophy}
    \end{subfigure}
    \hspace{0.005\textwidth}
    \begin{subfigure}[b]{0.32\textwidth}
        \centering
        \includegraphics[height=\kolfigheight,keepaspectratio]{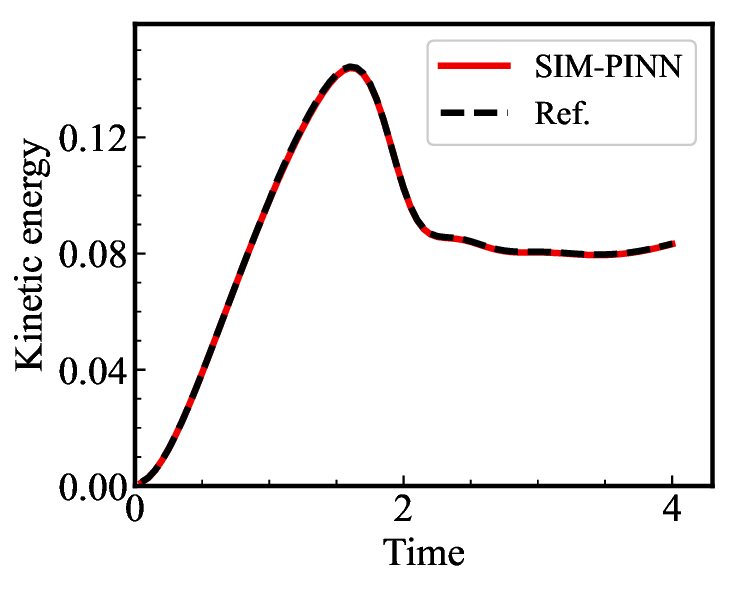}
        \caption{Kinetic energy}
        \label{fig:kolmogorov_ke}
    \end{subfigure}
    \hspace{0.005\textwidth}
    \begin{subfigure}[b]{0.32\textwidth}
        \centering
        \includegraphics[height=\kolfigheight,keepaspectratio]{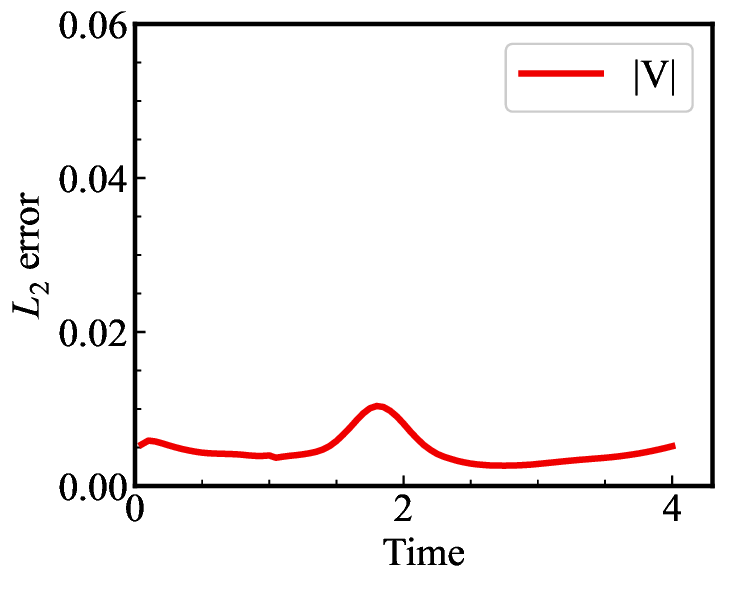}
        \caption{Relative $L_2$ error}
        \label{fig:kolmogorov_l2}
    \end{subfigure}
    \caption{Evolution of flow statistics and relative $L_2$ errors of the velocity magnitude for the two-dimensional Kolmogorov flow.}
    \label{fig:kolmogorov_stats}
\end{figure}

\section{Discussions}
\label{sec:discussions}
The main contribution of the proposed SIM-PINN lies in two aspects that distinguish it from existing PINNs for incompressible flow simulations.
The first is a projection-inspired, segregated and iterative framework that reorganizes the optimization of the incompressible NS equations.
By assigning the linearized momentum-prediction and pressure-correction subproblems to dedicated neural subsolvers, SIM-PINN eliminates the direct competition between the momentum and continuity residuals within a single composite loss function.
This segregated approach replaces the coupled nonlinear optimization problem with a sequence of decoupled and linearized subproblems.
The projection-based pressure correction and inner iterations then enforce mutual consistency between the separately trained velocity and pressure fields.
Upon convergence, this mechanism ensures that the solution satisfies the coupled NS equations.

The second aspect is the incorporation of the multi-stage neural network training into the proposed segregated and iterative framework.
While the multi-stage neural network for the pressure Poisson equation serves as a standalone solver, that for the momentum equation is seamlessly integrated into the iterative framework, with each stage corresponding to one sub-iteration.
The stage-by-stage refinement of the network directly drives the iterative convergence of the velocity field within the coupled system. 
The adoption of the multi-stage training mitigates the spectral biases inherent to neural networks and underpins the convergence of SIM-PINN in challenging flow problems.
Furthermore, the proposed SIM-PINN framework provides a unique approach to extending the multi-stage PINN to the nonlinear and coupled NS equations, which has been found challenging previously.

The computational cost and number of trainable parameters of SIM-PINN are higher than those of a conventional monolithic PINN with the same network architecture and optimization settings.
Rather than optimizing a single network for both the momentum and continuity equations, SIM-PINN performs several inner iterations at each time step.
At each iteration, a freshly initialized staged network is trained for each momentum equation, and a complete multi-stage training process is performed for the pressure Poisson equation.
However, when computational cost is measured by the effort required to obtain a stable and accurate solution, SIM-PINN is expected to be more computationally efficient than a conventional monolithic PINN, particularly for high-Reynolds-number turbulent flows.
This is because the latter typically plateaus at a non-negligible prediction error or even diverges.
This advantage is supported by the lid-driven cavity and unsteady Kolmogorov flow results presented in Section~\ref{sec:results}, together with the cost- and parameter-count-matched comparisons for the lid-driven cavity flow presented in~\ref{app:cost_matched_comparison}.

The present work focuses on comparing SIM-PINN against a well-configured monolithic PINN baseline to isolate the contribution of the proposed framework.
While comparisons with other recent PINN variants would provide additional insights, such comparisons are complicated by differences in training strategy and network architectures.
Many recent PINN studies address complementary aspects of PINN training, such as  adaptive loss balancing~\cite{wang2021understanding,wang2022when,mcclenny2023self}, exact boundary-condition enforcement~\cite{sukumar2022exact}, and gradient enhancement~\cite{yu2022gradient}.
These approaches are compatible with the proposed SIM-PINN framework and can, in principle, be incorporated into its dedicated neural subsolvers.
Investigating such combinations remains an important direction for future work.

\section{Conclusions}
\label{sec:conclusions}
This paper presents a segregated, iterative, and multi-stage physics-informed neural network\,(SIM-PINN) framework for solving the incompressible Navier--Stokes equations. 
The proposed method addresses the optimization challenges and accuracy limitations inherent in conventional monolithic PINN formulations by integrating the projection approach of classical low-Mach number CFD with the multi-stage neural network architecture.
By decomposing the governing equations into linearized momentum-prediction and pressure-correction sub-problems, which are subsequently coupled through inner iterations, SIM-PINN effectively mitigates the ill-conditioning of the original fully-coupled system.
Furthermore, a multi-stage neural network training strategy incorporating frequency-guided embeddings and adaptive correction scaling is introduced for each decoupled equation to progressively reduce unresolved residuals at successive stages.

Numerical experiments are conducted to validate the proposed framework against the conventional monolithic PINN and reference numerical solutions.
In the steady incompressible lid-driven cavity flow at $Re=100$, segregated frozen tests demonstrate that SIM-PINN substantially reduces the residuals for both the pressure-correction Poisson equation and the linearized momentum equations. 
For the relatively high-Reynolds-number cavity flow at $Re=2000$, SIM-PINN achieves significantly better agreement with the reference solution than the conventional PINN, particularly in resolving the flow features within the near-wall regions and corner vortices. 
In the two-dimensional unsteady Kolmogorov flow, SIM-PINN achieves deep convergence of momentum and divergence residuals within 2--4 inner iterations per physical time step, whereas the conventional monolithic PINN fails to converge.
Moreover, SIM-PINN accurately captures the vorticity evolution from the initial forcing-dominated state through Kelvin--Helmholtz instability to complex multi-scale vortical structures.
The predicted enstrophy and kinetic energy closely match the reference solution throughout the entire time evolution.

\appendix

\section{Monolithic PINN for NS equations}
\label{app:PINN}

The baseline PINN is formulated within the same time-marching framework as SIM-PINN.
Unlike SIM-PINN, however, it does not adopt the segregated and iterative coupling framework or the frequency-guided multi-stage training strategy.
Instead, the velocity and pressure fields are optimized simultaneously at each time step by minimizing the residuals of the momentum and continuity equations. 
The momentum residual in the $i$-th coordinate direction is defined as
\begin{equation}
\mathcal{R}_{M_i}
=
\frac{u_i^{n+1}-u_i^{n}}{\Delta t}
+
u_j^{n+1}\frac{\partial u_i^{n+1}}{\partial x_j}
+
\frac{1}{\rho}\frac{\partial p}{\partial x_i}
-
\nu \frac{\partial^2 u_i^{n+1}}{\partial x_j \partial x_j}
-
\frac{1}{\rho}f_i,
\label{eq:pinn_momentum_residual}
\end{equation}

To enforce the prescribed velocity boundary conditions, the boundary residual for the velocity component in the $i$ direction is defined as
\begin{equation}
\mathcal{R}_{M_i,bc}(\mathbf{x}_b)
=
u_i^{n+1}(\mathbf{x}_b)-u_{i,bc}(\mathbf{x}_b),
\quad
\mathbf{x}_b\in X_{\partial\Omega},
\label{eq:pinn_boundary_residual}
\end{equation}
where $\partial\Omega$ and $u_{i,bc}$ denote the boundary of the computational domain and the prescribed velocity values on the boundary, respectively.

Meanwhile, the incompressibility constraint is enforced through the continuity residual, which is defined as
\begin{equation}
\mathcal{R}_{c}(\mathbf{x})
=
\frac{\partial u_j^{n+1}(\mathbf{x})}{\partial x_j}.
\label{eq:pinn_continuity_residual}
\end{equation}

Based on the momentum, boundary, and continuity residuals defined above, the total loss function of the monolithic PINN is written as
\begin{equation}
\mathcal{L}_{\mathrm{PINN}}(\theta)
=
\sum_{i=1}^{d}
\left[
\frac{1}{N_\Omega}
\sum_{j=1}^{N_{\Omega}}
\left\|
\mathcal{R}_{M_i}(\mathbf{x}^{j})
\right\|_2^2
+
\frac{1}{N_{\partial\Omega}}
\sum_{j=1}^{N_{\partial\Omega}}
\left\|
\mathcal{R}_{M_i,bc}(\mathbf{x}_b^j)
\right\|_2^2
\right]
+
\frac{1}{N_\Omega}
\sum_{j=1}^{N_{\Omega}}
\left\|
\mathcal{R}_{c}(\mathbf{x}^{j})
\right\|_2^2,
\label{eq:pinn_total_loss}
\end{equation}
where $d$ is the spatial dimension.

\section{Network architecture and training optimization parameters for SIM-PINN}
\label{app:network_parameters}

Table~\ref{tab:network_architecture} summarizes the network architecture and training optimization parameters common to all SIM-PINN simulations in this work. 
Case-specific parameters are detailed in their respective sections.

\begin{table}[htbp]
\centering
\caption{Network architecture and training optimization parameters for SIM-PINN.}
\label{tab:network_architecture}
\begin{tabular*}{0.75\linewidth}{@{}l@{\extracolsep{\fill}}l@{}}
\toprule
Parameter & Value \\
\midrule
Number of Fourier features in RFF & 64 \\
Hidden layers & 4 \\
Neurons per hidden layer & 128 \\
Hidden-layer activation & $\tanh$ \\
Output-layer activation & Linear \\
Linear-layer weight normalization & Yes \\
Weight initialization & Normal distribution \\
Optimizer & Adam + L-BFGS \\
First-stage learning rate & Adam: $1.0\times10^{-3}$, L-BFGS: $1.0$ \\
Subsequent-stage learning rate & Adam: $5.0\times10^{-4}$, L-BFGS: $0.5$ \\
Training epochs per stage & Adam\,(5000) + L-BFGS\,(5) \\
\bottomrule
\end{tabular*}
\end{table}


\section{Cost- and parameter-count-matched comparisons for the lid-driven cavity flow}
\label{app:cost_matched_comparison}

The purpose of this appendix is to isolate the effects of the increased computational cost and number of trainable parameters on the accuracy advantage of SIM-PINN.
Two additional experiments, denoted as PINN-T and PINN-P, respectively, are conducted for the lid-driven cavity flow at $Re=2000$ with the monolithic PINN.
In PINN-T, the number of optimization epochs at each pseudo-time step is increased such that the wall-clock time per pseudo-time step, measured in the same computing environment, is comparable to that of SIM-PINN.
Specifically, at each pseudo-time step, the Adam optimizer is applied for $25{,}000$ epochs, followed by $10$ epochs of the L-BFGS optimizer.
In PINN-P, the widths of the hidden-layers are increased from $[128,128,128,128,128]$ to $[316,315,315,315,315]$.
The resulting network contains approximately six times as many trainable parameters as the baseline network, matching the total parameter count of SIM-PINN.
The number of pseudo-time steps in these two cases is set to $300$, which is the same as in SIM-PINN.
All other settings, including the collocation points and pseudo-time step size, remain the same as those for the monolithic PINN reported in Section~\ref{sec:steady_simu_SIM}.

\begin{figure}[htbp] 
    \centering

    \begin{subfigure}  {0.325\textwidth}
        \centering
        \includegraphics[width=\textwidth]{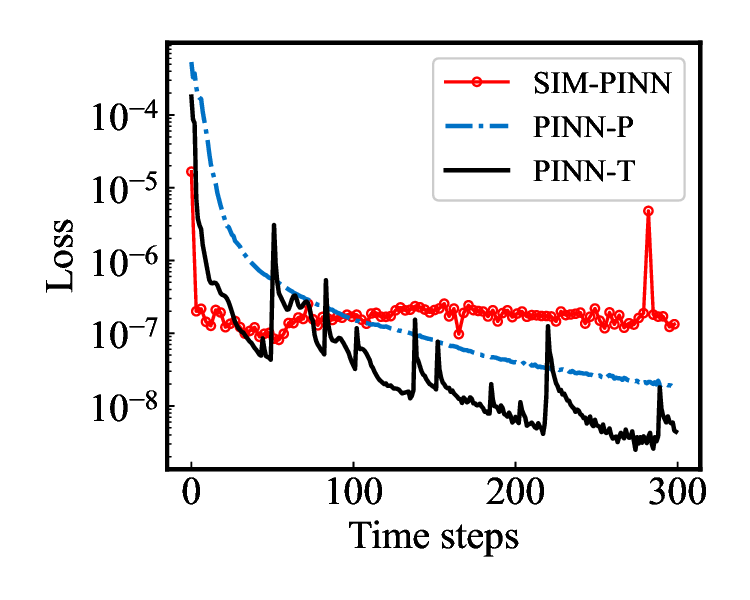}
        \caption{$x$-momentum equation}
        \label{fig:equal-mse-u}
    \end{subfigure}
    \begin{subfigure}  {0.325\textwidth}
        \centering
        \includegraphics[width=\textwidth]{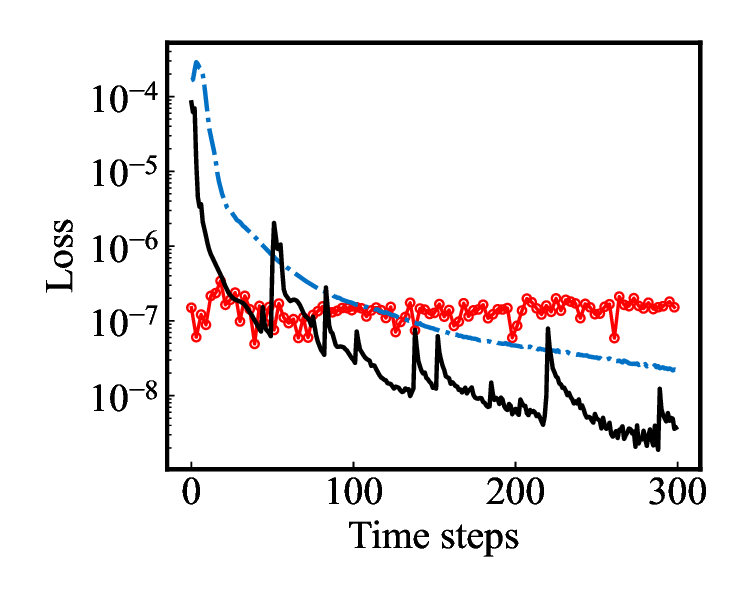}
        \caption{$y$-momentum equation}
        \label{fig:equal-mse-v}
    \end{subfigure}
    \begin{subfigure}  {0.325\textwidth}
        \centering
        \includegraphics[width=\textwidth]{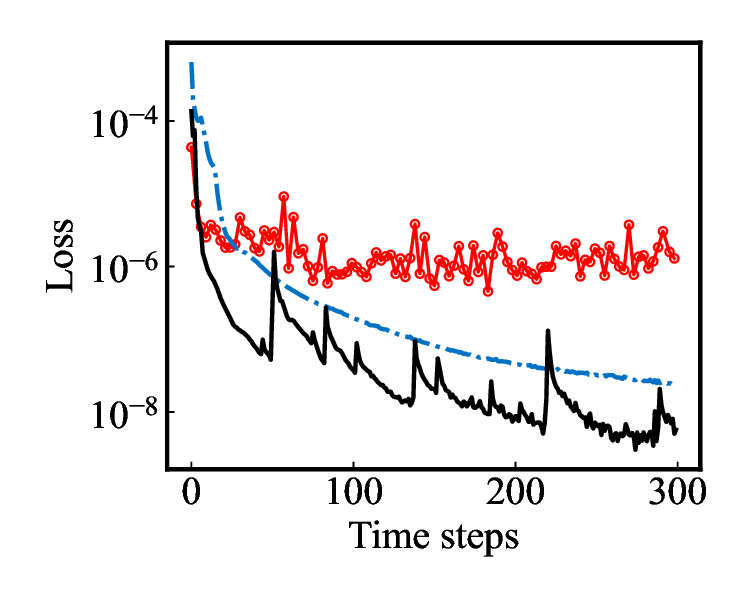}
        \caption{continuity equation}
        \label{fig:equal-mse-div}
    \end{subfigure}
    \caption{Comparison of the loss evolution for SIM-PINN, PINN-T, and PINN-P over $300$ pseudo-time steps for the lid-driven cavity flow at $Re=2000$.}
    \label{fig:cost-matched-mse-history}
\end{figure}

\begin{figure}[htbp] 
    \centering

    \begin{subfigure}[b]{0.48\textwidth}
        \centering
        \includegraphics[width=\textwidth]{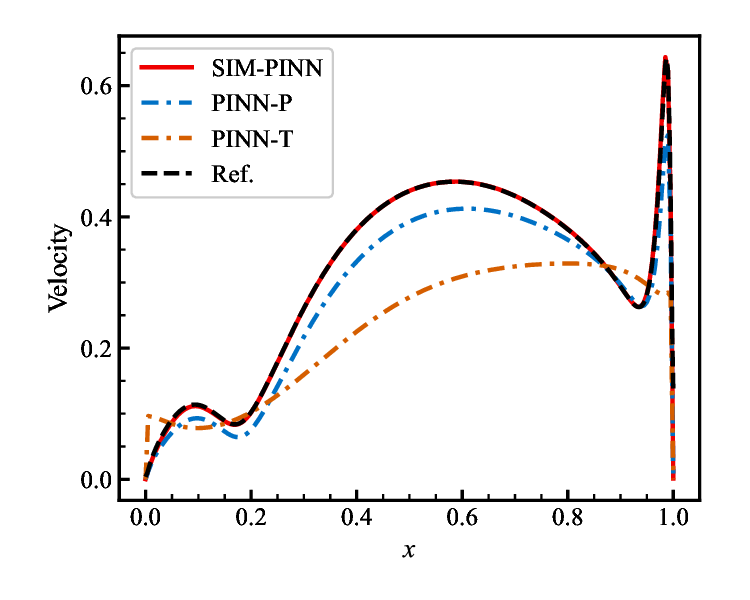}
        \caption{$y=0.95$}
        \label{fig:equal-sub-095}
    \end{subfigure}
    \hfill
    \begin{subfigure}[b]{0.48\textwidth}
        \centering
        \includegraphics[width=\textwidth]{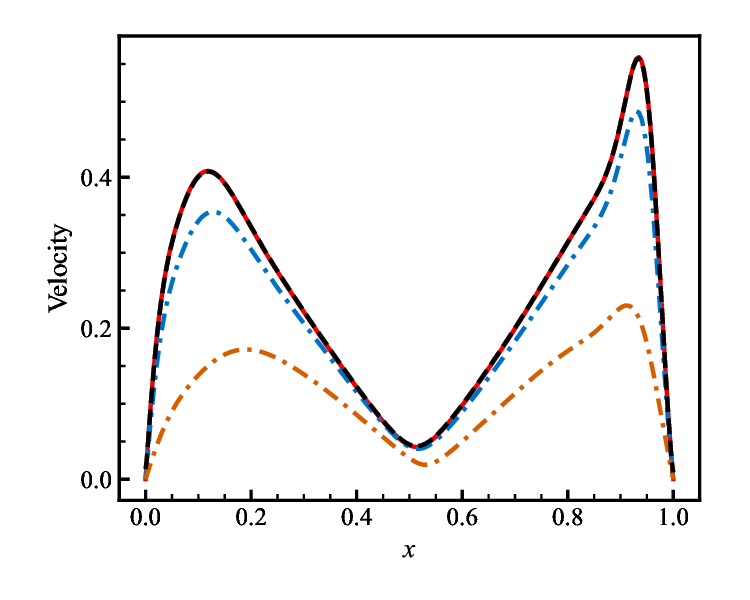}
        \caption{$y=0.5$}
        \label{fig:equal-sub-05}
    \end{subfigure}

    \begin{subfigure}[b]{0.48\textwidth}
        \centering
        \includegraphics[width=\textwidth]{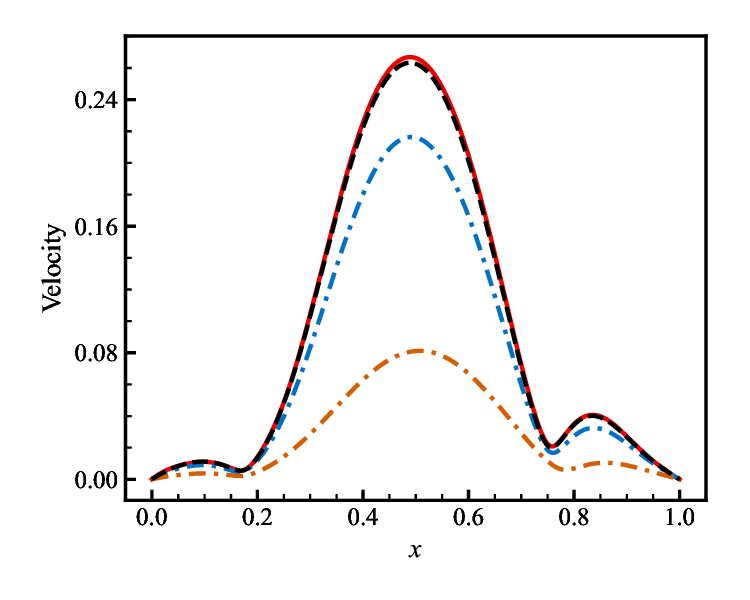}
        \caption{$y=0.05$}
        \label{fig:equal-sub-005}
    \end{subfigure}
    \hfill
    \begin{subfigure}[b]{0.48\textwidth}
        \centering
        \includegraphics[width=\textwidth]{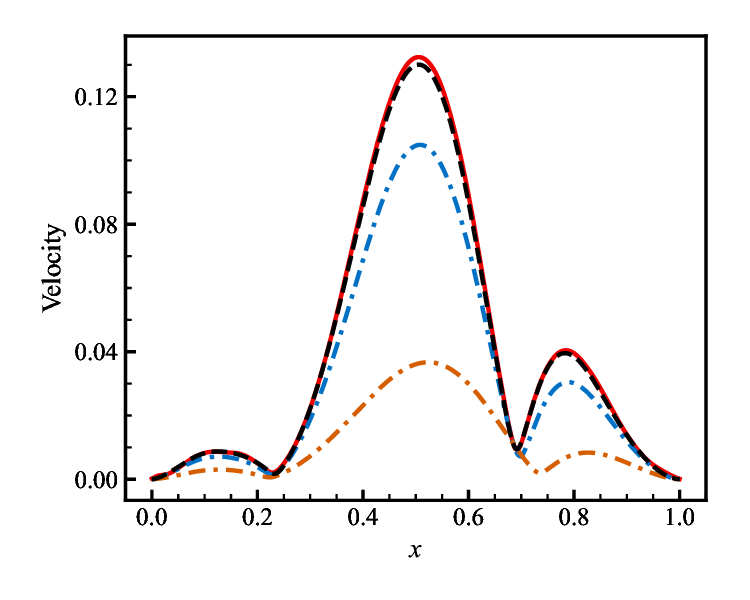}
        \caption{$y=0.02$}
        \label{fig:equal-sub-002}
    \end{subfigure}
    \caption{Comparison of velocity profiles along horizontal lines at different vertical locations obtained using SIM-PINN, PINN-T, and PINN-P for the lid-driven cavity flow at $Re=2000$.}
    \label{fig:cost-matched-velocity-profiles}
\end{figure}

Figure~\ref{fig:cost-matched-mse-history} compares the loss of the momentum and continuity equations among SIM-PINN, PINN-T, and PINN-P.
After the initial transient process, both PINN-T and PINN-P reach lower residual levels than SIM-PINN for both the momentum and continuity equation.
However, the lower residuals achieved with the monolithic PINN do not result in more accurate flow-field predictions.
As shown in Fig.~\ref{fig:cost-matched-velocity-profiles}, the velocity profiles predicted by PINN-T and PINN-P still deviate noticeably from the reference solution, whereas SIM-PINN maintains close agreement.
The mismatch between the training loss and solution accuracy in PINNs has also been reported previously~\cite{cao2023tsonn,Chiu2022}.
This has been attributed to the ill-conditioning of the corresponding optimization problem~\cite{krishnapriyan2021characterizing} by Cao and Zhang~\cite{cao2023tsonn}.
In the monolithic optimization framework, the training appears to settle into a suboptimal basin.
This issue is substantially alleviated in the proposed SIM-PINN framework, in which the decoupled and linearized subproblems are solved using dedicated neural networks and reconciled through projection-based iterations.

\section*{{CRediT} authorship contribution statement}

\textbf{Zhenyu Zhang}: Writing – review \& editing, Writing – original draft, Visualization, Software, Methodology, Investigation, Formal analysis, Data curation.
\textbf{Zhixiang Ling}: Writing – review \& editing, Investigation.
\textbf{Yinghao Huo}: Writing – review \& editing, Methodology, Investigation.
\textbf{Huayong Liu}: Writing – review \& editing, Investigation.
\textbf{Yunde Su}: Writing – review \& editing, Writing – original draft, Supervision, Methodology, Conceptualization, Funding acquisition, Project administration.

\section*{Declaration of competing interest}
The authors declare that they have no known competing financial interests or personal relationships that could have appeared to
influence the work reported in this paper.

\section*{Acknowledgments}
This work has been supported by Sichuan Science and Technology Program\,(No. 2026NSFSC0264), Sichuan University Interdisciplinary Innovation Fund\,(No. 1082204112L82), the Fundamental Research Funds for the Central Universities\,(No. 1082204112K92), and the National Numerical Windtunnel Program\,(No. NNW2025-ZT5-08).

\section*{Data Availability}
The datasets analysed during the current study are available from the corresponding author on reasonable request.

\bibliographystyle{elsarticle-num-names}
\bibliography{refs}

@ARTICLE{raissi2019physics,
  author  = {Raissi, M. and Perdikaris, P. and Karniadakis, G. E.},
  title   = {Physics-informed neural networks: {A} deep learning framework for solving forward and inverse problems involving nonlinear partial differential equations},
  journal = {J. Comput. Phys.},
  volume  = {378},
  year    = {2019},
  pages   = {686-707},
  publisher = {Elsevier}
}

@ARTICLE{raissi2020hidden,
  author  = {Raissi, M. and Yazdani, A. and Karniadakis, G. E.},
  title   = {Hidden fluid mechanics: {Learning} velocity and pressure fields from flow visualizations},
  journal = {Science},
  volume  = {367},
  number  = {6481},
  year    = {2020},
  pages   = {1026-1030},
  publisher = {American Association for the Advancement of Science}
}

@ARTICLE{baydin2018automatic,
  author  = {Baydin, A. G. and Pearlmutter, B. A. and Radul, A. A. and Siskind, J. M.},
  title   = {Automatic differentiation in machine learning: a survey},
  journal = {J. Mach. Learn. Res.},
  volume  = {18},
  number  = {153},
  year    = {2018},
  pages   = {1-43}
}

@ARTICLE{mcclenny2023self,
  author  = {McClenny, L. D. and Braga-Neto, U. M.},
  title   = {Self-adaptive physics-informed neural networks},
  journal = {J. Comput. Phys.},
  volume  = {474},
  year    = {2023},
  pages   = {111722},
  publisher = {Elsevier}
}

@ARTICLE{sukumar2022exact,
  author  = {Sukumar, N. and Srivastava, A.},
  title   = {Exact imposition of boundary conditions with distance functions in physics-informed deep neural networks},
  journal = {Comput. Methods Appl. Mech. Eng.},
  volume  = {389},
  year    = {2022},
  pages   = {114333},
  publisher = {Elsevier}
}

@ARTICLE{yu2022gradient,
  author  = {Yu, J. and Lu, L. and Meng, X. and Karniadakis, G. E.},
  title   = {Gradient-enhanced physics-informed neural networks for forward and inverse {PDE} problems},
  journal = {Comput. Methods Appl. Mech. Eng.},
  volume  = {393},
  year    = {2022},
  pages   = {114823},
  publisher = {Elsevier}
}

@ARTICLE{jin2021nsfnets,
  author  = {Jin, X. and Cai, S. and Li, H. and Karniadakis, G. E.},
  title   = {{NSFnets} ({Navier-Stokes} flow nets): {Physics-informed} neural networks for the incompressible {Navier-Stokes} equations},
  journal = {J. Comput. Phys.},
  volume  = {426},
  year    = {2021},
  pages   = {109951},
  publisher = {Elsevier}
}

@ARTICLE{wang2021understanding,
  author  = {Wang, S. and Teng, Y. and Perdikaris, P.},
  title   = {Understanding and mitigating gradient flow pathologies in physics-informed neural networks},
  journal = {SIAM J. Sci. Comput.},
  volume  = {43},
  number  = {5},
  year    = {2021},
  pages   = {A3055-A3081},
  publisher = {SIAM}
}

@ARTICLE{wang2022when,
  author  = {Wang, S. and Yu, X. and Perdikaris, P.},
  title   = {When and why {PINNs} fail to train: {A} neural tangent kernel perspective},
  journal = {J. Comput. Phys.},
  volume  = {449},
  year    = {2022},
  pages   = {110768},
  publisher = {Elsevier}
}

@INPROCEEDINGS{krishnapriyan2021characterizing,
  author  = {Krishnapriyan, A. and Gholami, A. and Zhe, S. and Kirby, R. and Mahoney, M. W.},
  title   = {Characterizing possible failure modes in physics-informed neural networks},
  booktitle = {Advances in Neural Information Processing Systems},
  volume  = {34},
  year    = {2021},
  pages   = {26548--26560}
}

@ARTICLE{wang2021eigenvector,
  author  = {Wang, S. and Wang, H. and Perdikaris, P.},
  title   = {On the eigenvector bias of {Fourier} feature networks: {From} regression to solving multi-scale {PDEs} with physics-informed neural networks},
  journal = {Comput. Methods Appl. Mech. Eng.},
  volume  = {384},
  year    = {2021},
  pages   = {113938},
  publisher = {Elsevier}
}

@ARTICLE{xu2019frequency,
  author  = {Xu, Z. Q. J. and Zhang, Y. and Luo, T. and Xiao, Y. and Ma, Z.},
  title   = {Frequency principle: {Fourier} analysis sheds light on deep neural networks},
  journal = {Commun. Comput. Phys.},
  volume  = {28},
  number  = {5},
  year    = {2020},
  pages   = {1746--1767}
}

@INPROCEEDINGS{rathore2024challenges,
  author  = {Rathore, P. and Lei, W. and Frangella, Z. and Lu, L. and Udell, M.},
  title   = {Challenges in training {PINNs}: {A} loss landscape perspective},
  booktitle = {Proceedings of the 41st International Conference on Machine Learning},
  series  = {Proceedings of Machine Learning Research},
  volume  = {235},
  year    = {2024},
  pages   = {42159--42191},
  publisher = {PMLR}
}

@ARTICLE{cao2025analysis,
  author  = {Cao, W. and Zhang, W.},
  title   = {An analysis and solution of ill-conditioning in physics-informed neural networks},
  journal = {J. Comput. Phys.},
  volume  = {520},
  year    = {2025},
  pages   = {113494},
  publisher = {Elsevier}
}

@ARTICLE{urban2025unveiling,
  author  = {Urb{\'a}n, J. F. and Stefanou, P. and Pons, J. A.},
  title   = {Unveiling the optimization process of physics informed neural networks: {How} accurate and competitive can {PINNs} be?},
  journal = {J. Comput. Phys.},
  volume  = {523},
  year    = {2025},
  pages   = {113656},
  publisher = {Elsevier}
}

@ARTICLE{song2025fenn,
  author  = {Song, J. and Cao, W. and Zhang, W.},
  title   = {{FENN}: {Feature-enhanced} neural network for solving partial differential equations involving fluid mechanics},
  journal = {J. Comput. Phys.},
  volume  = {542},
  year    = {2025},
  pages   = {114370},
  publisher = {Elsevier}
}

@ARTICLE{cao2025rans,
  author  = {Cao, W. and Shan, X. and Tang, S. and Ouyang, W. and Zhang, W.},
  title   = {Solving parametric high-{Reynolds}-number wall-bounded turbulence around airfoils governed by {Reynolds}-averaged {Navier--Stokes} equations using time-stepping-oriented neural network},
  journal = {Phys. Fluids},
  volume  = {37},
  number  = {1},
  year    = {2025},
  pages   = {015151},
  publisher = {AIP Publishing}
}

@ARTICLE{wang2024respecting,
  author  = {Wang, S. and Sankaran, S. and Perdikaris, P.},
  title   = {Respecting causality for training physics-informed neural networks},
  journal = {Comput. Methods Appl. Mech. Eng.},
  volume  = {421},
  year    = {2024},
  pages   = {116813},
  publisher = {Elsevier}
}

@ARTICLE{song2025matrix,
  author  = {Song, J. and Cao, W. and Zhang, W.},
  title   = {A matrix preconditioning framework for physics-informed neural networks based on the adjoint method},
  journal = {Phys. Fluids},
  volume  = {37},
  number  = {9},
  year    = {2025},
  pages   = {093614},
  publisher = {AIP Publishing}
}

@ARTICLE{chorin1968numerical,
  author  = {Chorin, A. J.},
  title   = {Numerical solution of the {Navier-Stokes} equations},
  journal = {Math. Comput.},
  volume  = {22},
  number  = {104},
  year    = {1968},
  pages   = {745-762}
}

@ARTICLE{wang2024multi,
  author  = {Wang, Y. and Lai, C. Y.},
  title   = {Multi-stage neural networks: {Function} approximator of machine precision},
  journal = {J. Comput. Phys.},
  volume  = {504},
  year    = {2024},
  pages   = {112865},
  publisher = {Elsevier}
}

@ARTICLE{cao2023tsonn,
  author  = {Cao, W. and Zhang, W.},
  title   = {{TSONN}: {Time-stepping-oriented} neural network for solving partial differential equations},
  journal = {arXiv preprint arXiv:2310.16491},
  year    = {2023}
}

@ARTICLE{kim1985application,
  author  = {Kim, J. and Moin, P.},
  title   = {Application of a fractional-step method to incompressible {Navier-Stokes} equations},
  journal = {J. Comput. Phys.},
  volume  = {59},
  number  = {2},
  year    = {1985},
  pages   = {308-323},
  publisher = {Elsevier}
}

@ARTICLE{brown2001accurate,
  author  = {Brown, D. L. and Cortez, R. and Minion, M. L.},
  title   = {Accurate projection methods for the incompressible {Navier--Stokes} equations},
  journal = {J. Comput. Phys.},
  volume  = {168},
  number  = {2},
  year    = {2001},
  pages   = {464-499},
  doi     = {10.1006/jcph.2001.6715},
  publisher = {Elsevier}
}

@ARTICLE{guermond2006overview,
  author  = {Guermond, J. L. and Minev, P. and Shen, J.},
  title   = {An overview of projection methods for incompressible flows},
  journal = {Comput. Methods Appl. Mech. Eng.},
  volume  = {195},
  number  = {44-47},
  year    = {2006},
  pages   = {6011-6045},
  publisher = {Elsevier}
}

@ARTICLE{jnini2025gauss,
  author  = {Jnini, A. and Vella, F. and Zeinhofer, M.},
  title   = {{Gauss--Newton} natural gradient descent for physics-informed computational fluid dynamics},
  journal = {Comput. Fluids},
  volume  = {307},
  year    = {2026},
  pages   = {106955},
  publisher = {Elsevier}
}

@ARTICLE{wang2025simulating,
  author  = {Wang, S. and Sankaran, S. and Fan, X. and Stinis, P. and Perdikaris, P.},
  title   = {Simulating three-dimensional turbulence with physics-informed neural networks},
  journal = {arXiv preprint arXiv:2507.08972},
  year    = {2025}
}

@ARTICLE{zhang2026physics,
  author  = {Zhang, W. and Suo, W. and Song, J. and Cao, W.},
  title   = {Physics-informed neural networks {(PINNs)} as intelligent computing technique for solving partial differential equations: {Limitation} and future prospects},
  journal = {Sci. China Phys. Mech. Astron.},
  volume  = {69},
  number  = {1},
  year    = {2026},
  pages   = {214602},
  publisher = {Springer}
}

@BOOK{Ferziger2002book,
  author  = {Ferziger, J. H. and Peri{\'c}, M.},
  title   = {Computational {Methods} for {Fluid} {Dynamics}},
  edition = {3},
  year    = {2002},
  doi     = {10.1007/978-3-642-56026-2},
  publisher = {Springer},
  address = {Berlin}
}

@ARTICLE{Desjardins2008,
  author  = {Desjardins, O. and Blanquart, G. and Balarac, G. and Pitsch, H.},
  title   = {High order conservative finite difference scheme for variable density low {Mach} number turbulent flows},
  journal = {J. Comput. Phys.},
  volume  = {227},
  number  = {15},
  year    = {2008},
  pages   = {7125-7159},
  doi     = {10.1016/j.jcp.2008.03.027},
  publisher = {Elsevier}
}

@ARTICLE{Su2018,
  author  = {Su, Y. and Kim, S. H.},
  title   = {An improved consistent, conservative, non-oscillatory and high order finite difference scheme for variable density low {Mach} number turbulent flow simulation},
  journal = {J. Comput. Phys.},
  volume  = {372},
  year    = {2018},
  pages   = {202-219},
  doi     = {10.1016/j.jcp.2018.06.021},
  publisher = {Elsevier}
}

@ARTICLE{zhu2024physics,
  author  = {Zhu, Y. and Kong, W. and Deng, J. and Bian, X.},
  title   = {Physics-informed neural networks for incompressible flows with moving boundaries},
  journal = {Phys. Fluids},
  volume  = {36},
  number  = {1},
  year    = {2024},
  pages   = {013617},
  publisher = {AIP Publishing}
}

@ARTICLE{patel2024turbulence,
  author  = {Patel, Y. and Mons, V. and Marquet, O. and Rigas, G.},
  title   = {Turbulence model augmented physics-informed neural networks for mean-flow reconstruction},
  journal = {Phys. Rev. Fluids},
  volume  = {9},
  number  = {3},
  year    = {2024},
  pages   = {034605},
  publisher = {APS}
}

@ARTICLE{cai2024physics,
  author  = {Cai, K. and Wang, J.},
  title   = {Physics-informed neural networks for solving incompressible {Navier--Stokes} equations in wind engineering},
  journal = {Phys. Fluids},
  volume  = {36},
  number  = {12},
  year    = {2024},
  pages   = {121303},
  publisher = {AIP Publishing}
}

@ARTICLE{xiao2024least,
  author  = {Xiao, Y. and Yang, L. M. and Shu, C. and Dong, H. and Du, Y. J. and Song, Y. X.},
  title   = {Least-square finite difference-based physics-informed neural network for steady incompressible flows},
  journal = {Comput. Math. Appl.},
  volume  = {175},
  year    = {2024},
  pages   = {33-48},
  publisher = {Elsevier}
}

@ARTICLE{gao2025grid,
  author  = {Gao, L. and Zhang, D. and Chen, Y. and Tu, Y. and Zhang, X. and Li, X.},
  title   = {Grid residual adaptive resampling for physics-informed neural networks to solve incompressible {Navier--Stokes} equations},
  journal = {J. Appl. Phys.},
  volume  = {137},
  number  = {15},
  year    = {2025},
  pages   = {154702},
  publisher = {AIP Publishing}
}

@ARTICLE{khademi2025physics,
  author  = {Khademi, A. and Dufour, S.},
  title   = {Physics-informed neural networks with trainable sinusoidal activation functions for approximating the solutions of the {Navier--Stokes} equations},
  journal = {Comput. Phys. Commun.},
  volume  = {314},
  year    = {2025},
  pages   = {109672},
  publisher = {Elsevier}
}

@ARTICLE{liu2025icpinn,
  author  = {Liu, Y. and Cai, L. and Chen, Y. and Chen, Q.},
  title   = {{ICPINN}: {Integral} conservation physics-informed neural networks based on adaptive activation functions for {3D} blood flow simulations},
  journal = {Comput. Phys. Commun.},
  volume  = {311},
  year    = {2025},
  pages   = {109569},
  publisher = {Elsevier}
}

@ARTICLE{wei2025ffv,
  author  = {Wei, C. and Fan, Y. and Wong, J. C. and Ooi, C. and Wang, H. and Chiu, P. H.},
  title   = {{FFV-PINN}: {A} fast physics-informed neural network with simplified finite volume discretization and residual correction},
  journal = {Comput. Methods Appl. Mech. Eng.},
  volume  = {444},
  year    = {2025},
  pages   = {118139},
  publisher = {Elsevier}
}

@ARTICLE{weller1998tensorial,
  author  = {Weller, H. G. and Tabor, G. and Jasak, H. and Fureby, C.},
  title   = {A tensorial approach to computational continuum mechanics using object-oriented techniques},
  journal = {Comput. Phys.},
  volume  = {12},
  number  = {6},
  year    = {1998},
  pages   = {620-631},
  publisher = {American Institute of Physics}
}

@ARTICLE{lucas2014spatiotemporal,
  author  = {Lucas, D. and Kerswell, R.},
  title   = {Spatiotemporal dynamics in two-dimensional {Kolmogorov} flow over large domains},
  journal = {J. Fluid Mech.},
  volume  = {750},
  year    = {2014},
  pages   = {518-554},
  publisher = {Cambridge University Press}
}

@ARTICLE{lucas2015recurrent,
  author  = {Lucas, D. and Kerswell, R. R.},
  title   = {Recurrent flow analysis in spatiotemporally chaotic 2-dimensional {Kolmogorov} flow},
  journal = {Phys. Fluids},
  volume  = {27},
  number  = {4},
  year    = {2015},
  publisher = {AIP Publishing}
}

@ARTICLE{chandler2013invariant,
  author  = {Chandler, G. J. and Kerswell, R. R.},
  title   = {Invariant recurrent solutions embedded in a turbulent two-dimensional {Kolmogorov} flow},
  journal = {J. Fluid Mech.},
  volume  = {722},
  year    = {2013},
  pages   = {554-595},
  publisher = {Cambridge University Press}
}

@ARTICLE{liu2025automatic,
  author  = {Liu, Z. and Liu, Y. and Yan, X. and Liu, W. and Nie, H. and Guo, S. and Zhang, C. A.},
  title   = {Automatic network structure discovery of physics informed neural networks via knowledge distillation},
  journal = {Nat. Commun.},
  volume  = {16},
  number  = {1},
  year    = {2025},
  pages   = {9558},
  publisher = {Nature Publishing Group UK London}
}

@ARTICLE{zhang2025physics,
  author  = {Zhang, Z. and Su, X. and Yuan, X.},
  title   = {Physics-informed neural networks with coordinate transformation to solve high {Reynolds} number boundary layer flows},
  journal = {J. Comput. Phys.},
  volume  = {541},
  year    = {2025},
  pages   = {114338},
  publisher = {Elsevier}
}

@ARTICLE{liu2024aspinn,
  author  = {Liu, Z. and Liu, Y. and Yan, X. and Liu, W. and Guo, S. and Zhang, C. A.},
  title   = {{AsPINN}: {Adaptive} symmetry-recomposition physics-informed neural networks},
  journal = {Comput. Methods Appl. Mech. Eng.},
  volume  = {432},
  year    = {2024},
  pages   = {117405},
  publisher = {Elsevier}
}

@ARTICLE{liu2023adaptive,
  author  = {Liu, Y. and Liu, W. and Yan, X. and Guo, S. and Zhang, C. A.},
  title   = {Adaptive transfer learning for {PINN}},
  journal = {J. Comput. Phys.},
  volume  = {490},
  year    = {2023},
  pages   = {112291},
  publisher = {Elsevier}
}

@ARTICLE{rahimi2007random,
  author  = {Rahimi, A. and Recht, B.},
  title   = {Random features for large-scale kernel machines},
  journal = {Adv. Neural Inf. Process. Syst.},
  volume  = {20},
  year    = {2007}
}

@ARTICLE{tancik2020fourier,
  author  = {Tancik, M. and Srinivasan, P. and Mildenhall, B. and Fridovich-Keil, S. and Raghavan, N. and Singhal, U. and Ramamoorthi, R. and Barron, J. and Ng, R.},
  title   = {{Fourier} features let networks learn high frequency functions in low dimensional domains},
  journal = {Adv. Neural Inf. Process. Syst.},
  volume  = {33},
  year    = {2020},
  pages   = {7537-7547}
}

@ARTICLE{kim2020front,
  author  = {Kim, S. H. and Su, Y.},
  title   = {Front propagation formulation for large eddy simulation of turbulent premixed flames},
  journal = {Combust. Flame},
  volume  = {220},
  year    = {2020},
  pages   = {439-454},
  publisher = {Elsevier}
}

@ARTICLE{su2023interaction,
  author  = {Su, Y. and Yang, Y.},
  title   = {Interaction between trefoil knotted flame and vortex},
  journal = {Phys. Rev. Fluids},
  volume  = {8},
  number  = {12},
  year    = {2023},
  pages   = {123201},
  publisher = {American Physical Society}
}

@BOOK{greenshields2023openfoam11,
  author    = {Greenshields, Christopher},
  title     = {{OpenFOAM v11 User Guide}},
  year      = {2023},
  publisher = {The OpenFOAM Foundation},
  address   = {London, UK},
  url       = {https://doc.cfd.direct/openfoam/user-guide-v11}
}

@article{Chiu2022,
title = {{CAN-PINN: A} fast physics-informed neural network based on coupled-automatic–numerical differentiation method},
journal = {Comput. Methods Appl. Mech. Eng.},
volume = {395},
pages = {114909},
year = {2022},
issn = {0045-7825},
author = {Pao-Hsiung Chiu and Jian Cheng Wong and Chinchun Ooi and My Ha Dao and Yew-Soon Ong}
}

@ARTICLE{wu2024variable,
  author  = {Wu, J. and Wu, Y. and Zhang, G. and Zhang, Y.},
  title   = {Variable linear transformation improved physics-informed neural networks to solve thin-layer flow problems},
  journal = {J. Comput. Phys.},
  volume  = {500},
  year    = {2024},
  pages   = {112761},
  doi     = {10.1016/j.jcp.2024.112761},
  publisher = {Elsevier}
}

@ARTICLE{li2025learning,
  author  = {Li, T. and Zou, Y. and Zou, S. and Chang, X. and Zhang, L. and Deng, X.},
  title   = {Learning to solve {PDEs} with finite volume-informed neural networks in a data-free approach},
  journal = {J. Comput. Phys.},
  volume  = {530},
  year    = {2025},
  pages   = {113919},
  doi     = {10.1016/j.jcp.2025.113919},
  publisher = {Elsevier}
}

\end{document}